\documentclass[version1996]{siggraph}
\usepackage{epsf}

\def\s{\sigma}
\def\sgn{\mathop{\rm sign}\nolimits}
\def\half{{\textstyle{{1}\over{2}}}}
\def\df{\partial}
\def\nn{\nonumber}

\def\Z{{\bf Z}}

\def\mod{\mathop{\mbox{mod}}\nolimits}

\def\ket#1{\bigl|#1\bigr>}
\def\bra#1{\bigl<#1\bigr|}
\def\vsp{\vspace{0mm}}

\newlength{\mywidth}\mywidth=2.3truein 
\epsfysize=\mywidth
\epsfxsize=\mywidth
\def\fref#1{fig.\ref{#1}}

\renewenvironment{figure}{\refstepcounter{figure}
\baselineskip=0.4\normalbaselineskip\footnotesize}
{\baselineskip=\normalbaselineskip}
\def\fignum{{\bf Fig.\arabic{figure}.\quad}}

\begin{document}

\title{String theory in Lorentz-invariant light cone gauge - II}
\author{Igor Nikitin, Lialia Nikitina\\
\\
{\it Fraunhofer Society, IMK.VE, 53754, St.Augustin, Germany}
}
\date{}
\maketitle

\quad

\vspace{-10mm}
\begin{center}
{\bf Abstract}
\end{center}

\baselineskip=0.4\normalbaselineskip\footnotesize

We perform a quantization of 4-dimensional Nambu-Goto theory of 
open string in light cone gauge, related in Lorentz-invariant way with the
world sheet. This allows to obtain a quantum theory without
anomalies in Lorentz group. We consider a special type of 
gauge fixing conditions, imposed in oscillator sector of the theory, 
which lead to a relatively simple Hamiltonian mechanics,
convenient for canonical quantization. We discuss 
the algebraic and geometric properties of this mechanics and
determine its mass spectrum for the states of spin singlet $S=0$.

\baselineskip=\normalbaselineskip\normalsize

\vsp\section{Introduction}

This paper continues the work \cite{partI}, further referred as Part~I,
devoted to Dirac's quantization of Nambu-Goto theory of open string,
formulated in the space-time of dimension d=4. The approach
consists in the introduction of light cone gauge with the gauge axis
related to a certain dynamical light-like vector in the theory. 
In this approach Lorentz group transforms the world sheet of the string
together with the gauge axis and is not followed by reparametrizations.
It is shown in Part~I, that under these conditions the quantum 
Lorentz group is free of anomalies. The details of the mechanics depend
on the choice of the dynamical vector to which the gauge axis
is attached, i.e. on the choice of {\it gauge fixing conditions}.
Generally the mechanics appears to be algebraically complex.
In this paper, Section~2 we consider gauge fixing conditions,
producing a resolvable system of polynomial equations.
We discuss the structure of the obtained solutions,
including their remaining discrete gauge symmetries.
In Section~3 we perform quantization of this mechanics
and determination of its mass spectrum for the states
of spin singlet $S=0$. In appendices we present the formulae
for the solutions of polynomial system, provide proofs of
the statements formulated throughout the paper
and give the details about numerical methods 
used for determination of mass spectrum.


\vsp\section{Classical mechanics}

\vsp\subsection{~\hspace{-3mm}\hbox{Lorentz-invariant abelian light cone gauge 
(lia-lcg)}}
We start from the canonical basis of string mechanics 
constructed in Part~I:

\begin{center}
$P_{\mu},Z_{\mu}$ +  infinite set  of  oscillators $a_{k},a_{k}^{*}$ 
+  the top ${\vec e_{i}},{\vec S}$,
\end{center}

\noindent
with Poisson brackets
\begin{eqnarray}
&&\{Z_{\mu},P_{\nu}\}=g_{\mu\nu},  \nn\\
&&\{a_{k},a_{n}^{*}\}=ik\delta_{kn},\ k,n\in\Z/\{0\},\label{Pb1}\\
&&\{S^{i},S^{j}\}=-\epsilon^{ijk}S^{k},  \
    \{S^{i},e^{j}_{n}\}=-\epsilon^{ijk}e^{k}_{n},\nn
\end{eqnarray}
and equivalent symplectic form
\begin{eqnarray}
&&~\hspace{-1.2cm}
\Omega=dP_{\mu}\wedge dZ_{\mu}+
 \sum_{k \neq 0}\frac{1}{ik}\; da_{k}^{*}\wedge da_{k}
+\half d{\vec e_{i}} \wedge d({\vec S}\times{\vec e_{i}}).
\label{symform}
\end{eqnarray}
The mechanics is restricted by 4 constraints of the 1st class:
$$\{\chi_{0},\chi_{i}\}=0\ ,\ 
\{\chi_{i},\chi_{j}\}=\epsilon_{ijk}\chi_{k},$$
which include mass shell condition and requirements
of the form ``spin of the top is equal to the spin of the string''\footnote{
In this paper the terms ``generator of rotations'', ``orbital moment''
and ``spin'' refer to the same variable. Finally it defines 
the spin of elementary particles, which will result from quantization.}: 
\begin{eqnarray}
&& \chi_{0}={\textstyle\frac{P^{2}}{2\pi}}-L_{0}=0,\quad 
L_{0}=\sum\limits_{n\neq0} a_{n}^{*}a_{n},\nn\\
&&\chi_{3}=S_{3}-A_{3}=0,\quad  
A_{3}=\sum\limits_{n\neq0} {\textstyle\frac{1}{n}}a_{n}^{*}a_{n},\nn\\ 
&&\chi_{+}= S_{+}-A_{+}=0,\quad  \chi_{-}=S_{-}-A_{-}=0,\nn\\
&&\chi_{\pm}=\chi_{1}\pm i\chi_{2},\quad  S_{\pm}=S_{1}\pm iS_{2},\nn\\
&& A_{-}=\sqrt{{\textstyle\frac{2\pi}{P^{2}}}}\sum_{k,n,k+n\neq0} 
{\textstyle\frac{1}{k}}a_{k}a_{n}a_{k+n}^{*},\label{Apol}\\
&& A_{+}=\sqrt{{\textstyle\frac{2\pi}{P^{2}}}}\sum_{k,n,k+n\neq0} 
{\textstyle\frac{1}{k}}a_{k}^{*}a_{n}^{*}a_{k+n},\nn
\end{eqnarray}
where $S_{i}=S^{k}e_{i}^{k}$ is a projection of ${\vec S}$ onto 
${\vec e}_{i}$.
Constraints generate the following transformations:
\begin{itemize}
\item $\chi_{0}$ generates phase shifts of oscillator variables\\
$E_{0}: a_{n}\to a_{n}e^{-in\tau}$ and translations of mean 
coordinate $Z_{\mu}\to Z_{\mu}+P_{\mu}\tau/\pi$;
\item $\chi_{3}$ generates phase shifts $R_{3}: a_{n}\to a_{n}e^{-i\alpha}$
and rotations of $\vec e_{1,2}$ about $\vec e_{3}$:
$\vec e_{i}\to R(\vec e_{3},-\alpha)\vec e_{i},\ i=1,2$;
\item $\chi_{1,2}$ generate rotations of basis $\vec e_{i}$
about axes $\vec e_{1,2}$ and certain non-linear transformations
of oscillator variables, which will be considered in more details 
in subsection~\ref{gribsec}.
\end{itemize}

The main difference of this mechanics from the standard light cone
gauge description consists in the fact that 
the gauge axis is included in the set of dynamical variables.
Its position in the center-of-mass frame is described by
unit norm 3-vector $\vec e_{3}$, while the gauge axis in Minkowski space 
is defined by light-like 4-vector
$n_{\mu}^{-}=(N_{\mu}^{0}-N_{\mu}^{i}e_{3}^{i})/\sqrt{2}$,
where $N_{\mu}^{\nu}$ is an orthonormal tetrad introduced in Part~I.
In familiar terms, $\chi_{0}$ generates the evolution
of the string, i.e. can be used as Hamiltonian; $\chi_{3}$ 
transforms the canonical variables of the theory in such a way
that observable string's coordinates and momenta are not changed,
therefore it can be interpreted as internal gauge 
transformation of the theory; $\chi_{1,2}$ change the orientation
of the gauge axis $\vec e_{3}$ and are followed by reparametrizations
of the string. Generators of Lorentz group defined 
in Part I produce ``rigid'' 
Lorentz transformations of the world sheet together
with the gauge axis, without change of its parametrization.
As a result, in quantum mechanics Lorentz group
was freed of anomalies. The central problem of this approach
is that the algebra of constraints $\chi_{i}$ in quantization 
acquires the same anomaly that earlier was in Lorentz group.
To solve this problem, we partially eliminate the gauge degrees 
of freedom in the classical theory, imposing additional
gauge fixing conditions to $\chi_{i}$. These conditions are 
equivalent to fixation of gauge axis $\vec e_{3}$ 
with respect to other dynamical vectors in the system.
As a result of this procedure, the set of constraints in the theory
will be reduced, and the remainder will be anomaly-free.

Before to proceed, we will consider one remarkable 
particular solution in string theory.

\vsp\paragraph*{Straight-line string} is a solution,
for which the string has a form of straight segment rotating about 
its middle at constant angular velocity in the center-of-mass
frame. This solution plays a special role in string theory \cite{slstring}: 
\begin{itemize}
\item it can be quantized at arbitrary number of dimensions $d\geq3$
without anomalies in Lorentz group;
\item it belongs to a border of classical Regge-plot 
$P^{2}/2\pi\geq S$, and in quantum theory corresponds 
to a leading Regge-trajectory $P^{2}/2\pi=S$;
\item it describes satisfactory the spectrum of light mesons 
and gives an opportunity to consider their radiative transitions.
\end{itemize}

In light cone gauge the straight-line string 
is represented by one-modal solution, if the gauge axis $\vec e_{3}$
is directed along the spin $\vec S$, orthogonally to the plane, 
where the straight-line string is rotating:
\begin{eqnarray}
&&~\hspace{-1cm}
a_{n}=0,\ n\neq1,\quad S_{\pm}=0,\quad
{\textstyle\frac{P^{2}}{2\pi}}=S=S_{3}=|a_{1}|^{2}.\label{1mode}
\end{eqnarray}
There is a gauge-equivalent one-modal solution with exited $(a_{-1})$-mode
and $\vec e_{3}$ opposite to $\vec S$. Other directions of gauge axis 
for the straight-line string give infinitely-modal solutions 
(particular example see in Part~I). 
Later we will use the solution (\ref{1mode}) to study the structure 
of general theory in its vicinity. 

\vsp\paragraph*{Additional gauge fixing conditions} we impose have a form
\begin{eqnarray}
&&a_{s}+a_{-s}=0,\quad a_{s}^{*}+a_{-s}^{*}=0\label{gauge}
\end{eqnarray}
for some $s>0$. It's easy to check that straight-line 
string solution (\ref{1mode}) satisfies these conditions
at $s>1$. Later, in subsection \ref{gribsec},
we will show that conditions (\ref{gauge})
can be imposed on any solution of string theory.

The gauge fixing conditions (\ref{gauge}) are preserved by
transformations $R_{3}$. They are not preserved by $E_{0}$,
however, there is a remainder of $E_{0}$-symmetry,
discrete transformation 
$$D_{2s}:\ a_{n}\to a_{n}e^{-in\pi/s}$$
preserving (\ref{gauge}): $a_{s}+a_{-s}\to-a_{s}-a_{-s}$.
As a result, $R_{3},D_{2s}$-symmetries are present in the theory 
after gauge fixation.

Substituting (\ref{gauge}) to the symplectic form (\ref{symform}),
we see that $a_{\pm s}$-terms cancel each other:
${\textstyle{1\over is}}(da_{s}^{*}\wedge da_{s}-da_{-s}^{*}\wedge da_{-s})=0$.
To study this property in more detail, we introduce the variables 
$q_{1}=\mbox{Re}(a_{s}+a_{-s})/2,\ q_{2}=-\mbox{Im}(a_{s}+a_{-s})/2,\ 
p_{1}=\mbox{Im}(a_{s}-a_{-s})/2,\ p_{2}=\mbox{Re}(a_{s}-a_{-s})/2$, 
and rewrite $a_{\pm s}$-terms of the symplectic form
as ${\textstyle{4\over s}}(dp_{1}\wedge dq_{1}-dp_{2}\wedge dq_{2})$.
Gauge fixing conditions (\ref{gauge}) are rewritten to $q_{1}=q_{2}=0$,
now we see that the variables $p_{1,2}$, canonically conjugated 
to $q_{1,2}$, drop out from the symplectic form. 
$p_{1,2}$ should be expressed in terms of other dynamical variables
from $\chi$-constraints, and independently on how complex
these expressions could be, the symplectic structure 
of the mechanics remains to be simple. This exclusive property
follows from the fact that two gauge fixing conditions (\ref{gauge})
are in involution with each other: $\{a_{s}+a_{-s},a_{s}^{*}+a_{-s}^{*}\}=0$,
i.e. generate {\it abelian} group of transformations.

Then we see that contribution of $a_{\pm s}$ to $A_{3}$ vanishes similarly:
${\textstyle{1\over s}}(|a_{s}|^{2}-|a_{-s}|^{2})=0$,
as a result, $a_{\pm s}$-terms also drop out from $\chi_{3}=S_{3}-A_{3}$.
This result follows from the fact that gauge fixing conditions (\ref{gauge}) 
are preserved by transformation $R_{3}$ and therefore are 
in involution with $\chi_{3}$.

The gauge fixing conditions (\ref{gauge}) 
are not in involution with $\chi_{0}$, 
and $a_{\pm s}$-terms in $L_{0}$ do not vanish: 
$|a_{s}|^{2}+|a_{-s}|^{2}=2|a_{s}|^{2}$. Poisson brackets of (\ref{gauge})
with $\chi_{\pm}$ also do not vanish. We conclude that (\ref{gauge}) 
are gauges for $(\chi_{\pm},\chi_{0})$ and $a_{\pm s}$ should be 
determined from these three constraints.


\vsp\subsection{Algebraic properties of lia-lcg}

Isolating contribution of $a_{\pm s}$-oscillators in (\ref{Apol})
and using the relation $a_{-s}=-a_{s}$, we have

\begin{eqnarray}
&&a_{s}^{2}d+\half a_{s}a_{s}^{*}d^{*}+a_{s}f+a_{s}^{*}g
+\Sigma_{-}-\sqrt{{\textstyle{P^{2}\over2\pi}}}S_{-}=0,\label{maineqs}\\
&&a_{s}^{*2}d^{*}+\half a_{s}a_{s}^{*}d+a_{s}^{*}f^{*}+a_{s}g^{*}
+\Sigma_{+}-\sqrt{{\textstyle{P^{2}\over2\pi}}}S_{+}=0,\nn\\
&&{\textstyle{P^{2}\over2\pi}}=L_{0}^{(s)}+2a_{s}a_{s}^{*},\nn
\end{eqnarray}
where
\begin{eqnarray}
&& d=d_{+}-d_{-},\ f=f_{+}-f_{-},\ g=g_{-}-g_{+},\label{defelem}\\
&&d_{+}=a_{2s}^{*}/s,\ d_{-}=a_{-2s}^{*}/s,\
L_{0}^{(s)}={\sum}'a_{k}^{*}a_{k}\nn\\
&&g_{-}={\sum}'
{\textstyle{1\over k}}a_{k}a_{s-k},\ 
g_{+}={\sum}'
{\textstyle{1\over k}}a_{k}a_{-s-k},\nn\\
&&f_{+}={\sum}'
\left({\textstyle{1\over s}}+{\textstyle{1\over k}}\right)
a_{k}a_{s+k}^{*},\ 
f_{-}={\sum}'
\left(-{\textstyle{1\over s}}+{\textstyle{1\over k}}\right)
a_{k}a_{-s+k}^{*},\nn\\
&&\Sigma_{-}={\sum}'
{\textstyle{1\over k}}a_{k}a_{n}a_{k+n}^{*},\
\Sigma_{+}={\sum}'
{\textstyle{1\over k}}a_{k}^{*}a_{n}^{*}a_{k+n},\
\Sigma_{-}^{*}=\Sigma_{+},\nn
\end{eqnarray}
Here in the sums ${\sum}'$ terms with $a_{0}^{(*)}$ and
$a_{\pm s}^{(*)}$ are excluded. Introducing denotations
\begin{eqnarray}
&&\lambda=\sqrt{{\textstyle{P^{2}\over2\pi}}},\
n_{s}=(\lambda^{2}-L_{0}^{(s)})/2,\nn\\
&&k=\Sigma_{-}+\half n_{s}d^{*}-\lambda S_{-},\nn
\end{eqnarray}
we can treat (\ref{maineqs}) as overdetermined polynomial system
for $(a_{s},a_{s}^{*})$ at given values of coefficients
$(d,f,g,k,n_{s})$:
\begin{eqnarray}
&&a_{s}^{2}d+a_{s}f+a_{s}^{*}g+k=0,\label{maineqs1}\\
&&a_{s}^{*2}d^{*}+a_{s}^{*}f^{*}+a_{s}g^{*}+k^{*}=0,\nn\\
&&a_{s}a_{s}^{*}-n_{s}=0.\nn
\end{eqnarray}
To solve this system we find its 
Groebner basis \cite{ideals,Mathematica}: 
\begin{eqnarray}
&&Gr[(\ref{maineqs1}),\{a_{s},a_{s}^{*}\}]=
\{H,a_{s}^{*}G+F,a_{s}^{*}\tilde G+\tilde F,...\},\label{groeb}
\end{eqnarray}
where $H,G,F,...$ are polynomials of coefficients,
given in Appendix~1. Groebner basis has the same set of roots
as original system (\ref{maineqs1}) and possesses a structure convenient for
their sequential determination. The first equation $H=0$
is a condition of consistency of the original system
formulated in terms of its coefficients. The next equations
give the solution of the system as rational expressions 
$$a_{s}^{*}=-F/G=-\tilde F/\tilde G=...$$

\begin{figure}\label{asexpr}
\begin{center}
~\epsfxsize=3.5cm\epsfysize=3.5cm\epsffile{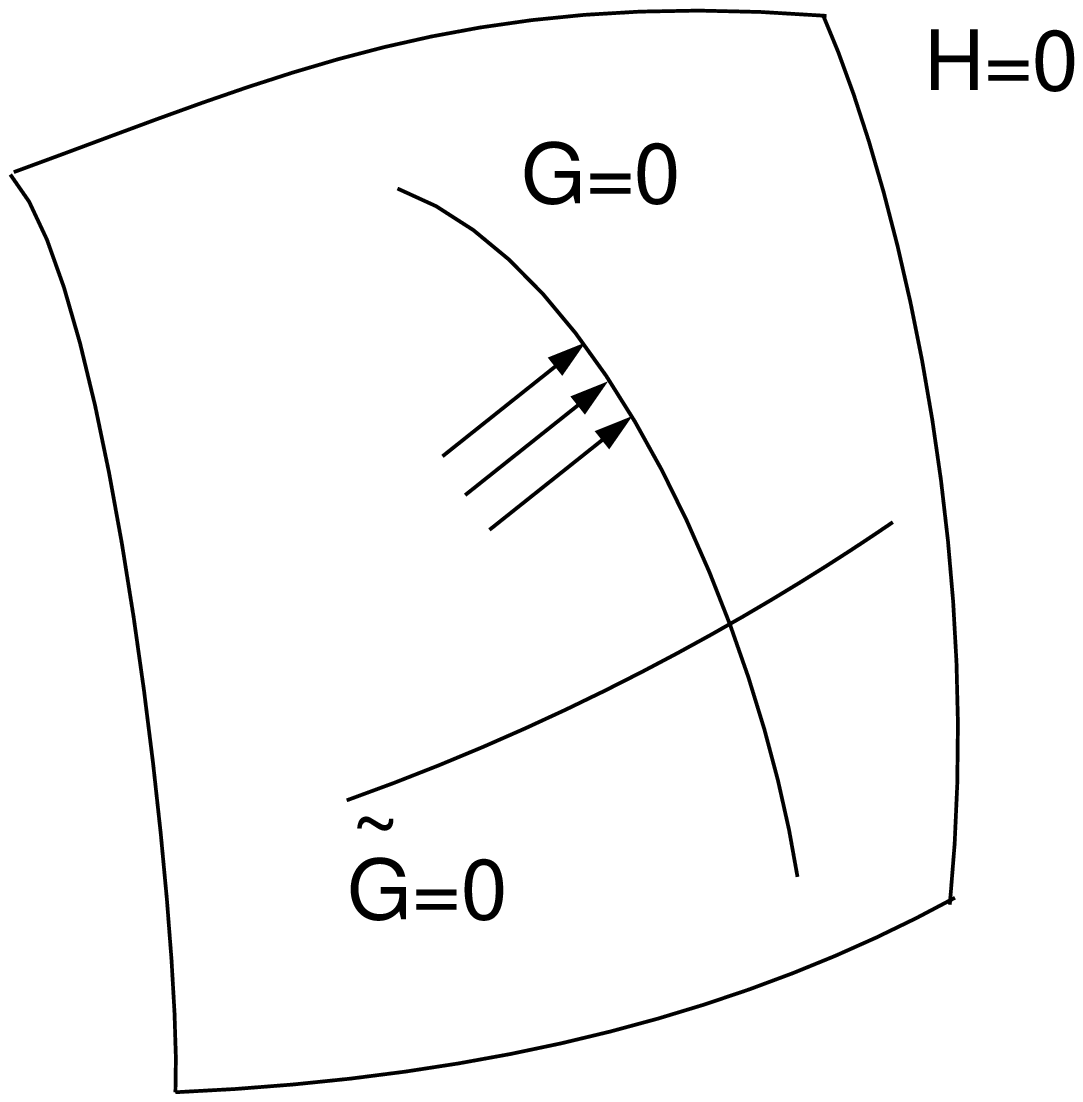}

\fignum Resolution of singularities in $a_{s}^{*}$.
\end{center}
\end{figure}

Different expressions for $a_{s}^{*}$ coincide on the surface $H=0$,
in the regions where the denominators do not vanish. 
In the points where the denominator in one formula vanishes, e.g. $G=0$,
we can use the other formula $a_{s}^{*}=-\tilde F/\tilde G$, provided 
that $\tilde G\neq0$. 
Therefore, in such points both numerator and denominator
of $a_{s}^{*}=-F/G$ must vanish simultaneously and resolution of this 
ambiguity gives finite value $(-\tilde F/\tilde G)$. 
At the end of Groebner basis the polynomials of degree $>1$  
in $a_{s}^{*}$ are present. Those which do not contain $a_{s}$,
describe solutions for the exceptional degenerate cases when
all denominators of the linear formulae vanish: $G=\tilde G=...=0$.
Those which are linear in $a_{s}$ formally can be used to find $a_{s}$.
The original equations are also appended at the end
of Groebner basis, meaning that all solutions must be
checked by substitution to the original system.

The presence of complex conjugation in (\ref{maineqs1}) creates
two general problems: (i) the described procedure treats 
$(a_{s},a_{s}^{*})$ 
as independent variables $(a_{s},b_{s})\in \bar C^{2}$, it
does not guarantee that $a_{s}$ will be complex conjugated to $b_{s}$.
However, the symmetry of original system with respect to
the complex conjugation implies that $(b_{s}^{*},a_{s}^{*})$
is also a solution of (\ref{maineqs1}), possibly the other one.
We are interested only by the case $(b_{s}^{*},a_{s}^{*})=(a_{s},b_{s})$.
(ii) theorems \cite{ideals} in the case of non-holomorphic polynomial systems  
do not guarantee that all solutions of $H=0$ will be continued
to full solutions of (\ref{maineqs1}), obstacles are the equations
of type $z^{*}z=-1$. To resolve both problems we substitute
the expression $a_{s}^{*}=-F/G$ and its complex conjugate
$a_{s}=-F^{*}/G^{*}$ to the original equations and explicitly
check that they are satisfied on the surface $H=0$
(after putting all terms over a common denominator
we check that $H$ is a polynomial divisor of the numerator).
As a result, we obtain the solutions $(a_{s},a_{s}^{*})$,
which are correctly transformed by complex conjugation
and are defined everywhere except of zero-measure 
set of degenerate cases described above.

$H=0$, the condition of consistency of the system (\ref{maineqs1}),
is the constraint, remaining after imposition of two gauge fixing
conditions (\ref{gauge}) to three initial constraints 
$\chi_{0},\chi_{\pm}$. This condition is real-valued: $H^{*}=H$.
Being used as Hamiltonian, it generates 
correct string evolution, consisting of shifts $\s\to\s+\tau$
and such reparametrizations that keep gauge fixing conditions (\ref{gauge}) 
permanently satisfied. Hamiltonian is in involution with the constraint
$\chi_{3}$, i.e. is preserved by transformation
$R_{3}: a_{n}\to a_{n}e^{-i\alpha},\ S_{-}\to S_{-}e^{-i\alpha}$.
Hamiltonian is also preserved by transformation
$D_{2s}: a_{n}\to a_{n}e^{-in\pi/s}$. Using the explicit
representation of $a_{s}$ from Appendix~1, one can easily 
verify that the properties of $a_{s}$
with respect to these transformations are the same as
before gauge fixation: 
\begin{eqnarray}
&&R_{3}(a_{s})=a_{s}e^{-i\alpha},\ D_{2s}(a_{s})=-a_{s}.\label{astransf}
\end{eqnarray}
It's important to remark that 
these symmetries are defined by polynomial structure of the expressions:
the symmetries implemented as phase rotations of oscillator variables
impose certain conditions on the powers of oscillator variables
in monomials. As a result, $R_{3},D_{2s}$-symmetries 
can be preserved in quantum theory as well. 

Generally $H$ is 8th degree polynomial in $\lambda$, in special
case $S_{\pm}=0$ it degenerates to 4th degree polynomial in $\lambda^{2}$.
Using implicit function theorems, one can locally rewrite 
the constraint $H=0$ in the form of mass-shell condition: 
$P^{2}-\Phi=0$, where the function $\Phi$ globally has multiple branches. 
Because at $s>1$ straight-line string (\ref{1mode}) satisfies 
gauge fixing conditions (\ref{gauge}), one branch of $\Phi$ passes through 
the straight-line string. Further we study in details the structure 
of solutions in the vicinity of straight-line string.

\vsp\subsection{Solutions in the vicinity of straight-line string}

\vspace{2mm}
\noindent{\bf Lemma~1:} at $s=2$ in the vicinity of straight-line string
the system (\ref{maineqs}) has unique solution, 
representable as $C^{\infty}$-smooth analytical function 
of coefficients of the system.

\noindent
[Proof of the lemma can be found in Appendix~2.]

\vspace{2mm}
Let's fix $s=2$. On the straight-line string solution (\ref{1mode})
for the coefficients of the system (\ref{maineqs})
we have $f_{\pm}=d_{\pm}=\Sigma_{\pm}=S_{\pm}=0$,
in $g$ the term $g_{+}=0$, while $g_{-}=a_{1}^{2}$.
Let's pull off $a_{1}$-contribution from $g$-terms:
$g'_{-}=g_{-}-a_{1}^{2}$, $g'=g'_{-}-g_{+}$.
Due to the lemma above we can represent
$a_{2}$ as Taylor's expansion with respect to 
these coefficients. Further we will study a part of this
expansion, which do not contain $S_{\pm}$-terms.
The case $S_{\pm}=0$ possesses convenient simplifications, 
particularly, the first two equations in (\ref{maineqs}) 
define a closed system for $(a_{2},a_{2}^{*})$ and 
decouple from the third one, which gives direct definition of $P^{2}$:
\begin{eqnarray}
&&{P^{2}\over2\pi}=L_{0}^{(2)}+2|a_{2}|^{2}.\nn
\end{eqnarray}

\vspace{2mm}
\noindent{\bf Lemma~2:} ($S_{\pm}=0$)-terms of $a_{2}$ expansion 
in the vicinity of straight-line string have a form

\begin{eqnarray}
&&a_{2}|_{S_{\pm}=0}=
\sum\limits_{n\geq1}{P_{n}a_{1}^{2}\over|a_{1}|^{4n}},\label{a2cl}
\end{eqnarray}
where $P_{n}$ are polynomials of $a_{1},\Sigma_{-},d,f,g'$ 
and their conjugates. The polynomials possess the following properties:
transformation $D_{4}: a_{n}\to a_{n}i^{-n}$ preserves 
the monomials of $P_{n}$, transformation 
$R_{3}: a_{n}\to a_{n}e^{-i\alpha}$ multiplies the monomials 
to $e^{i\alpha}$;
each monomial in $P_{n}$ contains, counting powers, $n$ variables 
from the group ($\Sigma_{-},d,f,g'$ and their conjugates) and
$2(n-1)$ variables from the group ($a_{1},a_{1}^{*}$);
each monomial contains at least one variable $\Sigma_{\pm}$;
numbers of variables from $(\Sigma_{\pm})$- and 
$(d,d^{*})$-groups are related: $n(\Sigma)=n(d)+1$.  

\vspace{2mm}
Explicit expressions for the first six polynomials $P_{n}$
can be found in Appendix~1. The expansion (\ref{a2cl})
has an obvious singularity at $a_{1}=0$. Later we will show
that in quantum theory this singularity can be removed
by a particular choice of quantum ordering.

\vsp\subsection{Geometric properties of lia-lcg}
\label{gribsec}
In this subsection we describe in more details the transformations, 
generated by constraints $\chi_{0},\chi_{i}$. At first, we introduce
several definitions.

Let's consider in 3D space: smooth closed curve $\vec Q(\s)$ 
with marked point $O$ 
and unit norm vector $\vec e_{3}$. Let's introduce variables
\begin{eqnarray}
&&~\hspace{-10mm}
{\vec a}_{n}={\textstyle{{1}\over{2\sqrt{2\pi}}}}
\oint d\vec Q(\s)\cdot\nn\\
&&\cdot\exp\left[
{\textstyle{{2\pi in}\over{L_{t}}}}
\left(L(\s)-(\vec Q(\s)-\vec Q(0))\vec e_{3}\right)\right],
\label{oint}
\end{eqnarray}
where $L(\s)$ is a length of arc between points $O$
and $\vec Q(\s)$ along the curve, $L_{t}$ is total length of the curve.
Two properties obviously follow from the definition:
${\vec a}_{-n}={\vec a}_{n}^{*},\ {\vec a}_{0}=0.$ 
Let's decompose vectors ${\vec a}_{n}$ into the components, parallel and
orthogonal to $\vec e_{3}$: 
${\vec a}_{n}=a_{n3}\vec e_{3}+{\vec a}_{n\perp}$,
and denote real and imaginary parts of ${\vec a}_{n\perp}$
as ${\vec q}_{n\perp}$ and ${\vec p}_{n\perp}$: 
${\vec a}_{n\perp}={\vec q}_{n\perp}+i{\vec p}_{n\perp}$. 
Let's fix some $n=s>0$ and write ${\vec q}_{s\perp}={\vec q}$ 
and ${\vec p}_{s\perp}={\vec p}$.
Functions ${\vec q}(\vec e_{3}),{\vec p}(\vec e_{3})$ 
define smooth vector fields on unit sphere of $\vec e_{3}$
(tangent to the sphere). Due to topological ``hedgehog'' theorem, 
these fields have singular points on the sphere, 
where corresponding field vanishes, e.g. ${\vec q}=0$.

\begin{figure}\label{qpfields}
\begin{center}
~\epsfxsize=6cm\epsfysize=3cm\epsffile{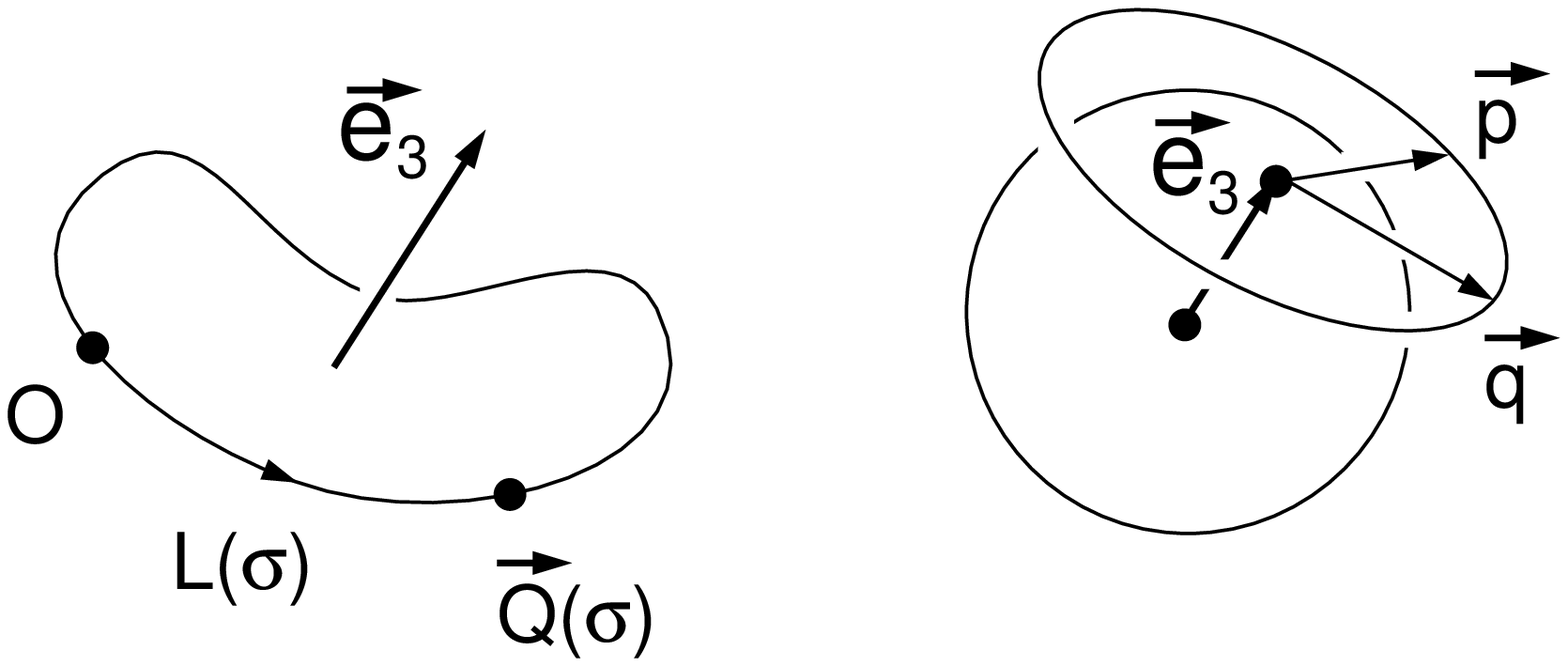}

\fignum Definition of vector fields $\vec q(\vec e_{3}),\vec p(\vec e_{3})$.

\end{center}
\end{figure}

\noindent{\it Note:}
the curve $\vec Q$ is a projection of {\it supporting curve},
introduced in Part I, into 3-dimensional subspace, orthogonal
to total momentum $P_{\mu}$. In terms of string dynamics,
it is a trajectory of string's end in the center-of-mass frame.
The length of this curve is equal to double mass of the string:
$L_{t}=2\sqrt{P^{2}}$. Gauge axis $\vec e_{3}$ relates 
the following parametrization to this curve:
\begin{eqnarray}
&&\s={{{2\pi}\over{L_{t}}}}(L(\s)-Q_{3}(\s)+Q_{3}(0)),
\label{lcgdef}
\end{eqnarray}
where $Q_{3}=\vec Q\vec e_{3}$. Now we recognize in (\ref{oint}) 
Fourier modes of function $\vec a(\s)=\vec Q'(\s)$, 
where $\s$ is lcg-parameter (\ref{lcgdef}). This expression
is written in parametric-invariant form, as circulation integral.
Vectors $\vec a_{n\perp}$ are related with earlier introduced 
oscillator variables $a_{n}$ as follows:
\begin{eqnarray}
&&\vec a_{n\perp}=a_{n1}\vec e_{1}+a_{n2}\vec e_{2},\nn\\
&&a_{n1}=(a_{n}+a_{-n}^{*})/2,\ a_{n2}=i(a_{n}-a_{-n}^{*})/2.\nn
\end{eqnarray}

\vspace{2mm}
\noindent{\bf Lemma~3:} $\chi_{3}$ generates transformations, preserving
$\vec e_{3}$ and $\vec a_{n\perp}$; $\chi_{1,2}$ generate rotations of
$\vec e_{3}$ and associated changes of $\vec a_{n\perp}$
according to the formula (\ref{oint}). 
Gauge fixing conditions (\ref{gauge}) correspond to a singular point ${\vec q}=0$. 
Evolution, generated by $\chi_{0}$, is represented as phase rotations 
${\vec a}_{n\perp}\to{\vec a}_{n\perp}e^{-in\tau}$.
During this evolution vector $\vec\omega=\vec q\times\vec p$ is preserved,
and singular points ${\vec q}=0$ move along zero-level curves
of a function $F(\vec e_{3})=\vec\omega\vec e_{3}=0$.

\vspace{2mm}
This lemma implies that gauge fixing conditions (\ref{gauge})
can be imposed on any solution of string theory.
Namely, these conditions can be satisfied for any curve 
$\vec Q(\s)$, directing the light cone gauge axis $\vec e_{3}$
to a singular point $\vec q(\vec e_{3})=0$ 
of a vector field on the sphere, constructed in terms
of this curve. 

In general position the vector fields on the sphere
have even number of singular points, which is $\geq2$. 
Non-degenerate singular points of 2D vector fields are \cite{DNF}

\begin{figure}\label{singtypes}
\begin{center}
~\epsfysize=2cm\epsfxsize=8cm\epsffile{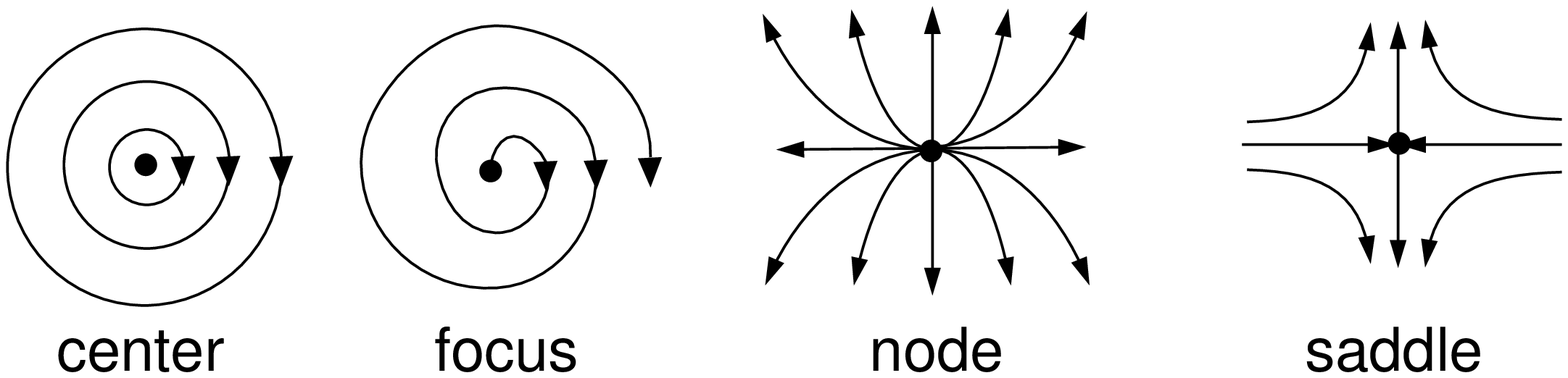}

\fignum Singular points of 2D vector field.

\end{center}
\end{figure}

\noindent
For each type 
{\it index} of singularity is defined as algebraic number of rotations
of vector $\vec q(\vec e_{3})$, when $\vec e_{3}$ passes around singular point
($>0$, if directions of rotations of $\vec q$ and $\vec e_{3}$ coincide;
$<0$ otherwise). For (center, node, focus) it is $+1$, for saddle $-1$.
The sum of all indices is equal to Euler characteristic of the surface
$V-E+F$: (num. of vertices) $-$ (num. of edges) $+$ (num. of faces)
for any tessellation of the surface into faces, which for the sphere
equals 2. Thus, generic vector field on a sphere has
2 singular points of index $+1$ and arbitrary number of self-compensating
pairs $(+1,-1)$. 


Presence of several singular points $\vec q(\vec e_{3})=0$ 
leads to the fact that an orbit of a gauge group generated 
by $\chi$-constraints intersects a surface of gauge fixing conditions 
(\ref{gauge})
in several points of the phase space. This phenomenon is also encountered
in the theory of non-abelian gauge fields, where it has been studied by
V.N.Gribov~\cite{Gribov}. We will call such points {\it Gribov's copies}. 

\begin{figure}\label{orbit}
\begin{center}
~\epsfxsize=4cm\epsfysize=2cm\epsffile{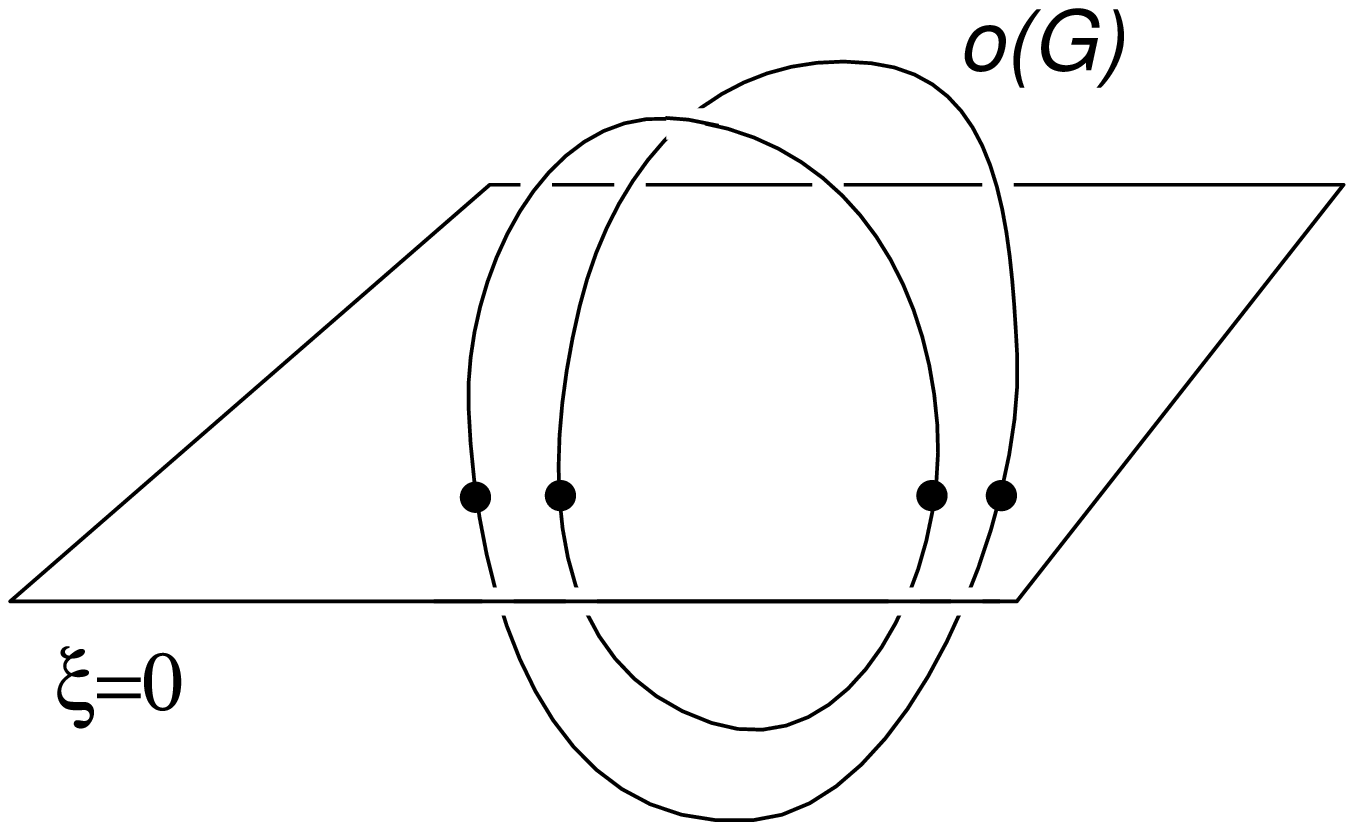}

\fignum Gribov's copies: the orbit of a gauge group $o(G)$
intersects the surface of gauge condition $\xi=0$
in several points.

\end{center}
\end{figure}

In our case the copies comprise different sets of oscillator variables 
$\{a_{n}\}$, reproducing the same curve $\vec Q(\s)$ 
(differently parameterized), i.e. physically correspond 
to the same string's motion. This property indicates 
remaining {\it discrete} gauge equivalence in the theory.
To identify the equivalent states, in classical theory 
the phase space should be factorized with respect to this symmetry. 
Analogous procedures can be applied in quantum mechanics, 
e.g. by constructing irreducible representations for operators 
possessing this discrete symmetry and formulating respective selection rules.

During the evolution Gribov's copies move along closed loops,
constructed by lifting of zero-level curves $F(\vec e_{3})=0$
to the space of oscillator variables. 
In more details, this construction is performed as follows:
vector $\vec e_{3}(\tau)$ moves along the contour $F(\vec e_{3})=0$ 
on the sphere, defined for a given value $s$, 
while for $n\neq s$ expression (\ref{oint}) defines 
time-dependent vector fields $\vec a_{n\perp}(\vec e_{3},\tau)$,
where we substitute $\vec e_{3}=\vec e_{3}(\tau)$.
Note that corresponding vector fields ${\vec q}_{n\perp},{\vec p}_{n\perp}$
generally do not vanish on these loops. The set of Gribov's copies possesses 
$D_{2s}$-symmetry, however in general case this symmetry 
does not exhaust the discrete gauge equivalence.


\vspace{2mm}
\noindent{\bf Lemma~4:} during the period of evolution $\Delta\tau=2\pi$
singular points $\vec q=0$ return to the
initial position $2s$ times. The lifting of the contour $F(\vec e_{3})=0$
to the space of oscillator variables has at least $2s$ Gribov's copies,
related by $D_{2s}$-symmetry.

\vspace{2mm}
Below we study the structure of Gribov's copies in the vicinity
of straight-line string solution (\ref{1mode}) and its 
$\vec e_{3}\to-\vec e_{3}$ companion. For these purpose 
we investigate the expansion of $\vec q(\vec e_{3})$ for this solution
in the vicinity of northern and southern poles of the sphere 
$\vec e_{3}=\pm \vec S/|\vec S|$.


\vspace{2mm}
\noindent{\bf Lemma~5:} $\vec q(\vec e_{3})$ for 
straight-line string solution 
at $s=1$ has no singularities in the vicinity of the poles;
at $s>1$ has in each pole a singularity, which does not move during 
the evolution, and at $s=2$ is a saddle point, at $s>2$ has degenerate type.


\vspace{2mm}
\noindent{\bf Lemma~6:} Solutions, close to (\ref{1mode}), 
have at $s=2$ a saddle point moving along a small loop 
in the vicinity of the northern pole. During the period of evolution 
$\Delta\tau=2\pi$ the saddle point performs 4 turns. 
Associated lifting has exactly 4 Gribov's copies, 
related by $D_{4}$-symmetry.

\vspace{2mm}
Analogous structure of Gribov's copies is established
for the companion solution near the southern pole.
Solutions close to straight-line string at $s=2$
should also have other singularities faraway of the poles
(saddle points have index $-2$, while the total index
of singular points on the sphere should be $+2$).
Further in quantum mechanics we will use the expansion (\ref{a2cl}) 
in the vicinity of solution (\ref{1mode}),
corresponding to the northern pole singularity.


\vsp\paragraph*{Summary of classical mechanics:}
the phase space $(Z,P,a_{n},a_{n}^{*},\vec S,\vec e_{i})$
possesses trivial topology. Hamiltonian $H$ is globally
defined in it as a polynomial function (Appendix~1). The following
two features can be considered as topological defects:
(i) the expression of mass squared $P^{2}=\Phi$ from the constraint $H=0$
is multi-valued; (ii) there is a discrete remainder of original gauge invariance,
which includes $D_{2s}$-symmetry, but generally is not exhausted by it.
In the vicinity of straight-line string solution (\ref{1mode}) at $s=2$
the branch of $P^{2}=\Phi$ is regular, and 
the discrete gauge invariance is exhausted by $D_{4}$-symmetry.

\vsp\section{Quantum mechanics}

\vsp\subsection{Canonical operators}

\begin{eqnarray}
&&[Z_{\mu},P_{\nu}]=-ig_{\mu\nu},\nn\\
&&[a_{k},a_{n}^{+}]=k\delta_{kn},\ k,n\neq0,\pm s,\label{canon}\\ 
&&[S^{i},S^{j}]=i\epsilon_{ijk}S^{k},\
[S^{i},e_{j}^{k}]=i\epsilon_{ikl}e_{j}^{l},\nn\\
&&[S_{i},S_{j}]=-i\epsilon_{ijk}S_{k},\
[S_{i},e_{j}^{k}]=-i\epsilon_{ijl}e_{l}^{k},\nn\\
&&[S^{i},S_{j}]=0,\ e_{i}^{k}e_{j}^{k}=\delta_{ij},\
S_{i}=e_{i}^{j}S^{j}.\nn 
\end{eqnarray}
The space of states is a direct product of three components:

\vsp\paragraph*{Space of functions $\Psi(P)$} 
with the definition $Z=-i\df/\df P$.

\vsp\paragraph*{Fock space} with a vacuum 

\begin{eqnarray}
&&a_{k}\ket{0}=0,\ k>0,\quad a_{k}^{+}\ket{0}=0,\ k<0\label{vacuum}
\end{eqnarray}
and states created from vacuum by operators

\begin{eqnarray}
&&\ket{\{n_{k}\}}=
\prod_{k>0,k\neq s}{{{1}\over{\sqrt{k^{n_{k}}n_{k}!}}}}
(a_{k}^{+})^{n_{k}}\cdot\nn\\
&&\cdot\prod_{k<0,k\neq-s}{{{1}\over{\sqrt{(-k)^{n_{k}}n_{k}!}}}}
(a_{k})^{n_{k}}
\ket{0}.\nn
\end{eqnarray}

For instance, we will write 
$\ket{1_{1}2_{-1}}={\textstyle{1\over\sqrt{2}}}a_{1}^{+}a_{-1}^{2}\ket{0}$
etc. So defined state vectors have positive norm. Occupation numbers
\begin{eqnarray}
&&n_{k}={{1}\over{|k|}}\;:a_{k}^{+}a_{k}:\;={{1}\over{|k|}}\;\cdot\;
  \left\{ \begin{array}{l}
  a_{k}^{+}a_{k},\;k>0\\
   a_{k}a_{k}^{+},\;k<0
  \end{array} \right.=0,1,2...\nn
\end{eqnarray}

\noindent{\it Note:} one-component oscillator variables $a_{k}$ 
used throughout this paper are related with commonly applied two-component
oscillators $a_{k1,2}$ by the formulae of subsection~\ref{gribsec},
which on quantum level become:
$a_{k}=a_{k1}-ia_{k2}$, $a_{k}^{+}=a_{-k1}+ia_{-k2}$.
The inverse formulae are $a_{k1}=(a_{k}+a_{-k}^{+})/2$,
$a_{k2}=i(a_{k}-a_{-k}^{+})/2$.
Here taking {\it two sets} of oscillator variables 
$a_{ki},i=1,2$ with $a_{ki}^{+}=a_{-ki}$ 
we construct {\it one set} without this property.
Usage of one-component oscillators simplifies the algebra.
It's easy to verify that definition of vacuum (\ref{vacuum})
is equivalent to a standard one $a_{ki}\ket{0}=0,k>0$
and the states $\ket{\{n_{k}\}}$ are the linear combinations
of $\prod_{k>0}(a_{k1}^{+})^{n_{k1}}(a_{k2}^{+})^{n_{k2}}\ket{0}$.

\vsp\paragraph*{Quantum top:} the space of states is formed by functions
$\Psi(e),\ e_{i}^{j}\in SO(3)$. Because the rotation group is not simply 
connected, $\pi_{1}(SO(3))=\Z_{2}$, two representations are possible:
single- and double-valued \cite{Weyl}. 
Spin is defined as differential operator 
$S^{i}=-i\epsilon_{ijk}e_{l}^{j}\partial/\partial e_{l}^{k}$,
while the projection of spin onto the coordinate system $\vec e_{i}$ is
$S_{i}=i\epsilon_{ijk}e^{l}_{j}\partial/\partial e^{l}_{k}$.
Operator $S^{i}$ generates the rotation of the coordinate system 
$\vec e_{i}$ in external space, while $S_{i}$ generates
the rotations about the axes $\vec e_{i}$. These transformations
act on different indices in $e_{j}^{k}$. They commute, therefore
$S^{3}$ and $S_{3}$ are simultaneously observable.
 
The basis in the space of states is formed by Wigner's ${\cal D}$-functions
$$\ket{SS_{3}S^{3}}={\cal D}^{S}_{S_{3}S^{3}}(e),\quad
 S_{3},S^{3}=-S,-S+1\; ...\; S,$$
where $S$ are integer for single-valued representation 
and half-integer for double-valued one. Further a constraint 
$S_{3}=A_{3}\in\Z$ will select only integer $S$-values\footnote{
However, using formal modifications $A_{3}\to A_{3}+1/2$,
classically vanishing after the recovery of Planck's constant,
it's possible to obtain half-integer spin as well.
The question on possibility of inclusion of half-integer 
$S$-values in this mechanics should be better investigated.
}. 


Operator $S^{2}=S^{i}S^{i}=S_{i}S_{i}$ has eigenvalues $S(S+1)$. 
The constraints, being functions of $S_{i}$ and oscillator variables,
commute with $S^{2}$, as a result, the determination of mass spectrum 
can be performed separately for each $S$ value. Particularly, 
spin singlet $S=0$ corresponds to state vectors
independent on $e_{i}^{j}$. On such states all $S^{i}=S_{i}=0$.

\vsp\paragraph*{Constraints} $\chi_{3}\ket{\Psi}=H\ket{\Psi}=0$
are imposed on the states in the spirit of Dirac's quantization
of constrained theories. Here $\chi_{3}=S_{3}-A_{3}^{(s)}$,

\begin{eqnarray}
&&A_{3}^{(s)}=
\sum_{k\neq 0,\pm s}{\textstyle{1\over k}}:a_{k}^{+}a_{k}:\;=
\sum_{k\neq 0,\pm s}\sgn k\cdot n_{k},\nn\\
&&L_{0}^{(s)}=
\sum_{k\neq 0,\pm s}:a_{k}^{+}a_{k}:\;=
\sum_{k\neq 0,\pm s}|k|\cdot n_{k}.\nn
\end{eqnarray}
In the next subsection we will construct two versions
of Hamiltonian $H$: polynomial of $\lambda=\sqrt{P^{2}/2\pi}$
and $P^{2}-\Phi$ with $\Phi$ is defined by Taylor's series (\ref{a2cl}).
For the operators entering in $H$: $d_{\pm},f_{\pm},g_{\pm},\Sigma_{\pm}$ 
and their conjugates we reserve a term {\it elementary operators}. 
They can be defined by their classical expressions (\ref{defelem}),
with only replacement of complex conjugation to Hermitian one.
These definitions have no ordering ambiguities. 
For the states with the finite number of filled modes the matrix 
elements of elementary operators are described by finite sums.

\vsp\paragraph*{Symmetries} $R_{3},E_{0},D_{2s}$ 
in quantum theory are generated by operators 
$$R_{3}=e^{i\chi_{3}\alpha},\ 
E_{0}=e^{iL_{0}^{(s)}\tau},\
D_{2s}=e^{iL_{0}^{(s)}\pi/s}.$$
It's convenient to introduce the notion of 
$(\Delta A_{3},\Delta L_{0})$-{\it charges}
for operators satisfying the relations:
$$[A_{3}^{(s)},op]=\Delta A_{3}\cdot op,\ 
[L_{0}^{(s)},op]=\Delta L_{0}\cdot op,$$ 
i.e. the operators, which increase or decrease the quantum numbers
$(A_{3}^{(s)},L_{0}^{(s)})$ by $(\Delta A_{3},\Delta L_{0})$ units:
$$op\ket{A_{3}^{(s)},L_{0}^{(s)}}=
\ket{A_{3}^{(s)}+\Delta A_{3},L_{0}^{(s)}+\Delta L_{0}}.$$
Particularly, $\Delta A_{3}(a_{n})=-1,\ \Delta L_{0}(a_{n})=-n$.
Hermitian conjugation of operators reverses their charges.
The elementary operators have definite 
$(\Delta A_{3},\Delta L_{0})$-charges, see Table~1.
The discrete symmetry $D_{2s}$ is also characterized by the charge 
$Q=L_{0}^{(s)}\mod 2s$, so that the subspaces with given $Q$
are eigenspaces $D_{2s}\ket{Q}=e^{i\pi Q/s}\ket{Q}$, and
$D_{2s}$-symmetric operators keep these subspaces invariant.

Because the symmetries of classical theory in our case are defined 
by its polynomial structure, they are preserved on the quantum level. 
Particularly, quantum Hamiltonian possesses $R_{3},D_{2s}$-symmetries,
as a result, its non-zero matrix elements form blocks, located on 
$A_{3}^{(s)}Q$-diagonal: $\bra{A_{3}^{(s)}Q}H\ket{A_{3}^{(s)}Q}$.

\vsp\subsection{Quantum Hamiltonian}

Hamiltonian $H$ can be defined directly by the expressions
of Appendix~1, where we only need to fix a proper ordering (Table~1).
As a result, we obtain a polynomial spectral problem \cite{matrix_pol} 
of the form $H(\lambda)\Psi_{\lambda}=0$, where
$H(\lambda)=\sum h_{n}\lambda^{n}$, the coefficients
$h_{n}$ are Hermitian operators and $\lambda=\sqrt{P^{2}/2\pi}$
is a spectral parameter. The problems of these kind are equivalent 
to a standard linear spectral problem in larger space:
$(M-\lambda)\tilde\Psi_{\lambda}=0$, where

\baselineskip=0.4\normalbaselineskip\footnotesize

$$M=\begin{array}{|ccccc|}\hline
&&&&\\
0&1&0&..&0\\
&&&&\\
0&0&1&..&0\\
&&&&\\
..&..&..&..&..\\
&&&&\\
0&0&0&..&1\\
&&&&\\
-\tilde h_{0}&-\tilde h_{1}&-\tilde h_{2}&..&-\tilde h_{l-1}\\
&&&&\\
\hline
\end{array}~,
$$

\baselineskip=\normalbaselineskip\normalsize

\noindent
$l$ is a degree of polynomial $H(\lambda)$ and
$\tilde h_{n}=h_{l}^{-1}h_{n}$, providing that $h_{l}^{-1}$ exists.
Using the regularizations based on the restriction of operators 
to finite-dimensional spaces $\ket{L_{0}^{(s)}\leq N}$ (described below), 
the problem is reduced to a spectral analysis of a large matrix, 
which can be performed with the aid of standard computational 
algorithms \cite{matrix_bible}. Due to the block structure of Hamiltonian, 
$\lambda$ spectrum can be defined separately in each $A_{3}^{(s)}Q$-block.

The following circumstance is rarely remarked.
Generally Hamiltonian $H$ can depend on $P^{2}$ non-linearly.
Classically one can resolve the constraint $H=0$
with respect to $P^{2}$, and obtain a locally equivalent
theory. On the quantum level these theories have principial
differences: the first one creates non-linear spectral problem,
while the second one gives the spectral problem $(P^{2}-\Phi)\Psi=0$, 
linear in terms of $P^{2}$. The last problem has relatively simple
structure of solutions: Hermitian operator $\Phi$ 
possesses complete orthonormal basis of eigenvectors
and real eigenvalues, representing the mass spectrum.
In the polynomial problem Hermitian operator $H(\lambda)$
at fixed $\lambda$ also has real eigenvalues
and complete orthonormal basis of eigenvectors:
$H(\lambda)\Psi_{\lambda,\mu}=\mu(\lambda)\Psi_{\lambda,\mu}$.
Here the eigenvalues and eigenvectors are the functions of $\lambda$. 
However, the vectors satisfying $H(\lambda)\Psi_{\lambda}=0$,
defined by the roots of the equation $\mu(\lambda)=0$, 
should not necessarily form a complete basis. For finite-dimensional
matrix systems the number of real roots of this equation
can be less than the dimension of the space, it also can be greater than
the dimension. Furthermore, the vectors $\Psi_{\lambda}=0$ for different
$\lambda$ generally are not orthogonal. 

The last property can be improved. In a particular case when
$\Psi_{\lambda}$ for the real roots $\lambda$ form
a complete basis, we can construct the matrix $Y=X\Lambda X^{-1}$,
where the columns of $X$ are vectors $\Psi_{\lambda}$
and $\Lambda=\mbox{diag}~\lambda$. It's easy to verify that
under these conditions $Y$ is a solution of matrix equation 
$\sum h_{n}Y^{n}=0$, where $Y^{n}$ stand on the right of $h_{n}$,
while $Y^{+}$ is a similar solution from the left: $\sum Y^{+n}h_{n}=0$.
In the classical limit both $Y$ and $Y^{+}$ correspond
to a real root $\lambda$ of classical equation 
$\sum h_{n}\lambda^{n}=0$. Operator $(Y+Y^{+})/2$ 
is Hermitian quantum analog of $\lambda$. It has orthogonal system 
of eigenvectors and can be used to define the mass spectrum.

The problem with the number of eigenvalues is more serious. 
If the classical equation $H(\lambda)=0$ has several real roots $\lambda_{i}$,
one can represent each root by a separate function on the phase space.
In quantum mechanics each root becomes Hermitian operator 
possessing a complete system of eigenvectors. However,
in the solutions of polynomial spectral problem
the corresponding eigensystems are mixed together and 
their separation appears to be non-trivial problem. 
Theory of matrix polynomials factorization \cite{matrix_pol} 
possesses sufficient potential to solve this problem.
Particularly, it introduces sign characteristics of eigenvalues,
which under certain conditions allow to separate them completely.
On the other hand, the resolution of constraint $H=0$ to
$P^{2}=\Phi$ and 


\def\figpage{
\begin{center}
Table 1: $L_{0}^{(2)}$-normally ordered elementary operators
$$\begin{array}{|c|cccccccccccccccc|}\hline
&a_{1}^{+}&(d_{-})^{+}&d_{+}&(f_{-})^{+}&f_{+}&(g_{-}')^{+}&g_{+}&
\Sigma_{-}&\Sigma_{+}&(g_{+})^{+}&g_{-}'&
(f_{+})^{+}&f_{-}&(d_{+})^{+}&d_{-}&a_{1}\\\hline
\Delta L_{0}^{(2)}&1&4&4&2&2&2&2&0&0&-2&-2&-2&-2&-4&-4&-1\\
\Delta A_{3}^{(2)}&1&-1&1&0&0&2&-2&-1&1&2&-2&0&0&-1&1&-1\\\hline
\end{array}
$$
\end{center}
}\twocolumn[
\vsp\paragraph*{Appendix 1:} the polynomials, entering
to the definition of Groebner's basis (\ref{groeb}) of the system 
(\ref{maineqs1}):

\vspace{5mm}
\baselineskip=0.4\normalbaselineskip\footnotesize

$H=
- {f^{*}}\,g\,{g^{*}}^2\,k^2 - f\,g^2\,{g^{*}}\,{k^{*}}^2 
- d\,{f^{*}}\,g\,k\,{k^{*}}^2 - {d^{*}}\,f\,{g^{*}}\,k^2\,{k^{*}} 
+ {d^{*}}\,{g^{*}}^2\,k^3 + d\,g^2\,{k^{*}}^3 
+ {d^{*}}\,f^2\,{f^{*}}\,{g^{*}}\,k\,n_{s} 
+ d\,f\,{f^{*}}^2\,g\,{k^{*}}\,n_{s}
+ 2\,{d^{*}}\,f^2\,g\,{g^{*}}\,{k^{*}}\,n_{s} 
+ 2\,d\,{f^{*}}^2\,g\,{g^{*}}\,k\,n_{s} 
- 3\,{d^{*}}\,f\,g\,{g^{*}}^2\,k\,n_{s} 
- 3\,d\,{f^{*}}\,g^2\,{g^{*}}\,{k^{*}}\,n_{s} 
- 2\,d\,{d^{*}}\,{f^{*}}\,{g^{*}}\,k^2\,n_{s} 
- 2\,d\,{d^{*}}\,f\,g\,{k^{*}}^2\,n_{s} 
- d^2\,{f^{*}}^3\,g\,n_{s}^2 
- {d^{*}}^2\,f^3\,{g^{*}}\,n_{s}^2  
+ d^2\,{d^{*}}\,{f^{*}}^2\,k\,n_{s}^2 
+ d\,{d^{*}}^2\,f^2\,{k^{*}}\,n_{s}^2  
+ 3\,d\,{d^{*}}^2\,f\,{g^{*}}\,k\,n_{s}^2 
+ 3\,d^2\,{d^{*}}\,{f^{*}}\,g\,{k^{*}}\,n_{s}^2  
+ f\,{f^{*}}\,g\,{g^{*}}\,k\,{k^{*}} 
+ g^2\,{g^{*}}^2\,k\,{k^{*}} 
+ d\,{d^{*}}\,k^2\,{k^{*}}^2  
- g^3\,{g^{*}}^3\,n_{s} 
+ 2\,f\,{f^{*}}\,g^2\,{g^{*}}^2\,n_{s} 
- f^2\,{f^{*}}^2\,g\,{g^{*}}\,n_{s}  
- d\,{d^{*}}\,f\,{f^{*}}\,k\,{k^{*}}\,n_{s} 
+ d\,{d^{*}}\,g\,{g^{*}}\,k\,{k^{*}}\,n_{s}  
- d\,{d^{*}}\,f\,{f^{*}}\,g\,{g^{*}}\,n_{s}^2 
+ 3\,d\,{d^{*}}\,g^2\,{g^{*}}^2\,n_{s}^2 
- 2\,d^2\,{d^{*}}^2\,k\,{k^{*}}\,n_{s}^2  
- 3\,d^2\,{d^{*}}^2\,g\,{g^{*}}\,n_{s}^3 
- d^2\,{d^{*}}^2\,f\,{f^{*}}\,n_{s}^3  
+ d^3\,{d^{*}}^3\,n_{s}^4 
$

\vspace{2mm}
$F=
-{f^{*}}^3\,g^2\,{g^{*}}\,k\,n_{s}
+ 2\,{d^{*}}\,{f^{*}}^2\,g\,{g^{*}}\,k^2\,n_{s} 
- {d^{*}}^2\,{f^{*}}\,{g^{*}}\,k^3\,n_{s} 
+ {f^{*}}^2\,g^3\,{g^{*}}\,{k^{*}}\,n_{s} 
- {d^{*}}^2\,g\,{g^{*}}\,k^2\,{k^{*}}\,n_{s} 
- {d^{*}}\,g^3\,{g^{*}}\,{k^{*}}^2\,n_{s} 
+ d\,{f^{*}}^4\,g^2\,n_{s}^2 
- {d^{*}}\,{f^{*}}\,g^3\,{g^{*}}^2\,n_{s}^2 
- 2\,d\,{d^{*}}\,{f^{*}}^3\,g\,k\,n_{s}^2 
- 2\,{d^{*}}^2\,f\,{f^{*}}\,g\,{g^{*}}\,k\,n_{s}^2 
+ {d^{*}}^2\,g^2\,{g^{*}}^2\,k\,n_{s}^2 
+ d\,{d^{*}}^2\,{f^{*}}^2\,k^2\,n_{s}^2 
+ 2\,{d^{*}}^3\,f\,{g^{*}}\,k^2\,n_{s}^2 
- 3\,d\,{d^{*}}\,{f^{*}}^2\,g^2\,{k^{*}}\,n_{s}^2 
+ 2\,{d^{*}}^2\,f\,g^2\,{g^{*}}\,{k^{*}}\, n_{s}^2 
+ 4\,d\,{d^{*}}^2\,{f^{*}}\,g\,k\,{k^{*}}\,n_{s}^2 
- d\,{d^{*}}^3\,k^2\,{k^{*}}\,n_{s}^2 
+ d\,{d^{*}}^2\,g^2\,{k^{*}}^2\,n_{s}^2 
+ 2\,d\,{d^{*}}^2\,f\,{f^{*}}^2\,g\,n_{s}^3 
- {d^{*}}^3\,f^2\,g\,{g^{*}}\,n_{s}^3 
+ 2\,d\,{d^{*}}^2\,{f^{*}}\,g^2\,{g^{*}}\,n_{s}^3 
- 2\,d\,{d^{*}}^3\,f\,{f^{*}}\,k\,n_{s}^3 
- 2\,d\,{d^{*}}^3\,g\,{g^{*}}\,k\,n_{s}^3 
- 2\,d\,{d^{*}}^3\,f\,g\,{k^{*}}\,n_{s}^3 
+ d\,{d^{*}}^4\,f^2\,n_{s}^4 
- d^2\,{d^{*}}^3\,{f^{*}}\,g\,n_{s}^4 
+ d^2\,{d^{*}}^4\,k\,n_{s}^4$

\vspace{2mm}
$G={f^{*}}^3\,g^2\,k\,{k^{*}} 
- 2\,{d^{*}}\,{f^{*}}^2\,g\,k^2\,{k^{*}} 
+ {d^{*}}^2\,{f^{*}}\,k^3\,{k^{*}} 
- {f^{*}}^2\,g^3\,{k^{*}}^2 
+ {d^{*}}^2\,g\,k^2\,{k^{*}}^2 
+ {d^{*}}\,g^3\,{k^{*}}^3 
- f\,{f^{*}}^4\,g^2\,n_{s} 
+ {f^{*}}^3\,g^3\,{g^{*}}\,n_{s} 
+ 2\,{d^{*}}\,f\,{f^{*}}^3\,g\,k\,n_{s} 
- 3\,{d^{*}}\,{f^{*}}^2\,g^2\,{g^{*}}\,k\,n_{s} 
- {d^{*}}^2\,f\,{f^{*}}^2\,k^2\,n_{s} 
+ 3\,{d^{*}}^2\,{f^{*}}\,g\,{g^{*}}\,k^2\,n_{s} 
- {d^{*}}^3\,{g^{*}}\,k^3\,n_{s} 
+ 3\,{d^{*}}\,f\,{f^{*}}^2\,g^2\,{k^{*}}\,n_{s} 
- 2\,{d^{*}}^2\,f\,{f^{*}}\,g\,k\,{k^{*}}\,n_{s} 
- {d^{*}}^3\,f\,k^2\,{k^{*}}\,n_{s} 
- 3\,{d^{*}}^2\,f\,g^2\,{k^{*}}^2\,n_{s} 
- 2\,{d^{*}}^2\,f^2\,{f^{*}}^2\,g\,n_{s}^2 
+ 2\,{d^{*}}^3\,f^2\,{f^{*}}\,k\,n_{s}^2 
+ 3\,{d^{*}}^3\,f^2\,g\,{k^{*}}\,n_{s}^2 
- {d^{*}}^4\,f^3\,n_{s}^3$

\baselineskip=\normalbaselineskip\normalsize

\vspace{5mm}
The polynomials, entering to the expansion (\ref{a2cl})
of variable $a_{2}|_{S_{\pm}=0}$ in the vicinity of straight-line
string solution (\ref{1mode}):

\vspace{5mm}
\baselineskip=0.4\normalbaselineskip\footnotesize

$P_{1}=-\Sigma_{+},$

\vspace{2mm}
$P_{2}={{a_{1}^{*}}}^2\,{f^{*}}\,{\Sigma_{-}} + 
  {{a_{1}}}^2\,{{g'}^{*}}\,{\Sigma_{+}},$

\vspace{2mm}
$2P_{3}=
-2\,{{a_{1}^{*}}}^4\,{f^{*}}\,{g'}\,{\Sigma_{-}} - 
  2\,{{a_{1}}}^2\,{{a_{1}^{*}}}^2\,{f^{*}}\,
   {{g'}^{*}}\,{\Sigma_{-}} - 
  2\,{{a_{1}^{*}}}^4\,{d^{*}}\,{{\Sigma_{-}}}^2 - 
  2\,{{a_{1}}}^2\,{{a_{1}^{*}}}^2\,f\,{f^{*}}\,
   {\Sigma_{+}} - 2\,{{a_{1}}}^4\,{{{g'}^{*}}}^2\,
   {\Sigma_{+}} - {{a_{1}}}^2\,{{a_{1}^{*}}}^2\,d\,
   {\Sigma_{-}}\,{\Sigma_{+}}$,
 
\vspace{2mm}
$2P_{4}=
2\,{{a_{1}}}^2\,{{a_{1}^{*}}}^4\,f\,{{f^{*}}}^2\,
   {\Sigma_{-}} + 2\,{{a_{1}^{*}}}^6\,{f^{*}}\,
   {{g'}}^2\,{\Sigma_{-}} + 
  2\,{{a_{1}}}^2\,{{a_{1}^{*}}}^4\,{f^{*}}\,
   {g'}\,{{g'}^{*}}\,{\Sigma_{-}} + 
  2\,{{a_{1}}}^4\,{{a_{1}^{*}}}^2\,{f^{*}}\,
   {{{g'}^{*}}}^2\,{\Sigma_{-}} + 
  {{a_{1}}}^2\,{{a_{1}^{*}}}^4\,d\,{f^{*}}\,
   {{\Sigma_{-}}}^2 + 4\,{{a_{1}^{*}}}^6\,{d^{*}}\,
   {g'}\,{{\Sigma_{-}}}^2 + 
  2\,{{a_{1}}}^2\,{{a_{1}^{*}}}^4\,{d^{*}}\,
   {{g'}^{*}}\,{{\Sigma_{-}}}^2 + 
  2\,{{a_{1}}}^2\,{{a_{1}^{*}}}^4\,f\,{f^{*}}\,
   {g'}\,{\Sigma_{+}} + 
  4\,{{a_{1}}}^4\,{{a_{1}^{*}}}^2\,f\,{f^{*}}\,
   {{g'}^{*}}\,{\Sigma_{+}} + 
  2\,{{a_{1}}}^6\,{{{g'}^{*}}}^3\,{\Sigma_{+}} + 
  4\,{{a_{1}}}^2\,{{a_{1}^{*}}}^4\,{d^{*}}\,f\,
   {\Sigma_{-}}\,{\Sigma_{+}} + 
  {{a_{1}}}^2\,{{a_{1}^{*}}}^4\,{d^{*}}\,{f^{*}}\,
   {\Sigma_{-}}\,{\Sigma_{+}} + 
  {{a_{1}}}^2\,{{a_{1}^{*}}}^4\,d\,{g'}\,
   {\Sigma_{-}}\,{\Sigma_{+}} + 
  2\,{{a_{1}}}^4\,{{a_{1}^{*}}}^2\,d\,{{g'}^{*}}\,
   {\Sigma_{-}}\,{\Sigma_{+}} + 
  {{a_{1}}}^4\,{{a_{1}^{*}}}^2\,d\,f\,{{\Sigma_{+}}}^2 + 
  2\,{{a_{1}}}^4\,{{a_{1}^{*}}}^2\,d\,{f^{*}}\,
   {{\Sigma_{+}}}^2$,
 
\vspace{2mm}
$4P_{5}=
-8\,{{a_{1}}}^2\,{{a_{1}^{*}}}^6\,f\,{{f^{*}}}^2\,
   {g'}\,{\Sigma_{-}} - 
  4\,{{a_{1}^{*}}}^8\,{f^{*}}\,{{g'}}^3\,
   {\Sigma_{-}} - 8\,{{a_{1}}}^4\,{{a_{1}^{*}}}^4\,f\,
   {{f^{*}}}^2\,{{g'}^{*}}\,{\Sigma_{-}} - 
  4\,{{a_{1}}}^2\,{{a_{1}^{*}}}^6\,{f^{*}}\,
   {{g'}}^2\,{{g'}^{*}}\,{\Sigma_{-}} - 
  4\,{{a_{1}}}^4\,{{a_{1}^{*}}}^4\,{f^{*}}\,
   {g'}\,{{{g'}^{*}}}^2\,{\Sigma_{-}} - 
  4\,{{a_{1}}}^6\,{{a_{1}^{*}}}^2\,{f^{*}}\,
   {{{g'}^{*}}}^3\,{\Sigma_{-}} - 
  12\,{{a_{1}}}^2\,{{a_{1}^{*}}}^6\,{d^{*}}\,f\,
   {f^{*}}\,{{\Sigma_{-}}}^2 - 
  2\,{{a_{1}}}^2\,{{a_{1}^{*}}}^6\,{d^{*}}\,
   {{f^{*}}}^2\,{{\Sigma_{-}}}^2 - 
  4\,{{a_{1}}}^2\,{{a_{1}^{*}}}^6\,d\,{f^{*}}\,
   {g'}\,{{\Sigma_{-}}}^2 - 
  12\,{{a_{1}^{*}}}^8\,{d^{*}}\,{{g'}}^2\,
   {{\Sigma_{-}}}^2 - 4\,{{a_{1}}}^4\,{{a_{1}^{*}}}^4\,d\,
   {f^{*}}\,{{g'}^{*}}\,{{\Sigma_{-}}}^2 - 
  8\,{{a_{1}}}^2\,{{a_{1}^{*}}}^6\,{d^{*}}\,
   {g'}\,{{g'}^{*}}\,{{\Sigma_{-}}}^2 - 
  4\,{{a_{1}}}^4\,{{a_{1}^{*}}}^4\,{d^{*}}\,
   {{{g'}^{*}}}^2\,{{\Sigma_{-}}}^2 - 
  2\,{{a_{1}}}^2\,{{a_{1}^{*}}}^6\,d\,{d^{*}}\,
   {{\Sigma_{-}}}^3 - 4\,{{a_{1}}}^4\,{{a_{1}^{*}}}^4\,f^2\,
   {{f^{*}}}^2\,{\Sigma_{+}} - 
  4\,{{a_{1}}}^2\,{{a_{1}^{*}}}^6\,f\,{f^{*}}\,
   {{g'}}^2\,{\Sigma_{+}} - 
  8\,{{a_{1}}}^4\,{{a_{1}^{*}}}^4\,f\,{f^{*}}\,
   {g'}\,{{g'}^{*}}\,{\Sigma_{+}} - 
  12\,{{a_{1}}}^6\,{{a_{1}^{*}}}^2\,f\,{f^{*}}\,
   {{{g'}^{*}}}^2\,{\Sigma_{+}} - 
  4\,{{a_{1}}}^8\,{{{g'}^{*}}}^4\,{\Sigma_{+}} - 
  8\,{{a_{1}}}^4\,{{a_{1}^{*}}}^4\,d\,f\,{f^{*}}\,
   {\Sigma_{-}}\,{\Sigma_{+}} - 
  8\,{{a_{1}}}^4\,{{a_{1}^{*}}}^4\,d\,{{f^{*}}}^2\,
   {\Sigma_{-}}\,{\Sigma_{+}} - 
  16\,{{a_{1}}}^2\,{{a_{1}^{*}}}^6\,{d^{*}}\,f\,
   {g'}\,{\Sigma_{-}}\,{\Sigma_{+}} - 
  4\,{{a_{1}}}^2\,{{a_{1}^{*}}}^6\,{d^{*}}\,
   {f^{*}}\,{g'}\,{\Sigma_{-}}\,{\Sigma_{+}} - 
  2\,{{a_{1}}}^2\,{{a_{1}^{*}}}^6\,d\,{{g'}}^2\,
   {\Sigma_{-}}\,{\Sigma_{+}} - 
  16\,{{a_{1}}}^4\,{{a_{1}^{*}}}^4\,{d^{*}}\,f\,
   {{g'}^{*}}\,{\Sigma_{-}}\,{\Sigma_{+}} - 
  4\,{{a_{1}}}^4\,{{a_{1}^{*}}}^4\,{d^{*}}\,
   {f^{*}}\,{{g'}^{*}}\,{\Sigma_{-}}\,{\Sigma_{+}} - 
  4\,{{a_{1}}}^4\,{{a_{1}^{*}}}^4\,d\,{g'}\,
   {{g'}^{*}}\,{\Sigma_{-}}\,{\Sigma_{+}} - 
  6\,{{a_{1}}}^6\,{{a_{1}^{*}}}^2\,d\,{{{g'}^{*}}}^2\,
   {\Sigma_{-}}\,{\Sigma_{+}} - 
  {{a_{1}}}^4\,{{a_{1}^{*}}}^4\,d^2\,{{\Sigma_{-}}}^2\,
   {\Sigma_{+}} - 4\,{{a_{1}}}^2\,{{a_{1}^{*}}}^6\,
   {{d^{*}}}^2\,{{\Sigma_{-}}}^2\,{\Sigma_{+}} - 
  4\,{{a_{1}}}^4\,{{a_{1}^{*}}}^4\,{d^{*}}\,f^2\,
   {{\Sigma_{+}}}^2 - 2\,{{a_{1}}}^4\,{{a_{1}^{*}}}^4\,
   {d^{*}}\,f\,{f^{*}}\,{{\Sigma_{+}}}^2 - 
  2\,{{a_{1}}}^4\,{{a_{1}^{*}}}^4\,d\,f\,{g'}\,
   {{\Sigma_{+}}}^2 - 4\,{{a_{1}}}^4\,{{a_{1}^{*}}}^4\,d\,
   {f^{*}}\,{g'}\,{{\Sigma_{+}}}^2 - 
  6\,{{a_{1}}}^6\,{{a_{1}^{*}}}^2\,d\,f\,{{g'}^{*}}\,
   {{\Sigma_{+}}}^2 - 12\,{{a_{1}}}^6\,{{a_{1}^{*}}}^2\,d\,
   {f^{*}}\,{{g'}^{*}}\,{{\Sigma_{+}}}^2 - 
  9\,{{a_{1}}}^4\,{{a_{1}^{*}}}^4\,d\,{d^{*}}\,
   {\Sigma_{-}}\,{{\Sigma_{+}}}^2 - 
  2\,{{a_{1}}}^6\,{{a_{1}^{*}}}^2\,d^2\,{{\Sigma_{+}}}^3$,
 
\vspace{2mm}
$4P_{6}=
4\,{{a_{1}}}^4\,{{a_{1}^{*}}}^6\,f^2\,{{f^{*}}}^3\,
   {\Sigma_{-}} + 12\,{{a_{1}}}^2\,{{a_{1}^{*}}}^8\,f\,
   {{f^{*}}}^2\,{{g'}}^2\,{\Sigma_{-}} + 
  4\,{{a_{1}^{*}}}^{10}\,{f^{*}}\,{{g'}}^4\,
   {\Sigma_{-}} + 16\,{{a_{1}}}^4\,{{a_{1}^{*}}}^6\,f\,
   {{f^{*}}}^2\,{g'}\,{{g'}^{*}}\,{\Sigma_{-}} + 
  4\,{{a_{1}}}^2\,{{a_{1}^{*}}}^8\,{f^{*}}\,
   {{g'}}^3\,{{g'}^{*}}\,{\Sigma_{-}} + 
  12\,{{a_{1}}}^6\,{{a_{1}^{*}}}^4\,f\,{{f^{*}}}^2\,
   {{{g'}^{*}}}^2\,{\Sigma_{-}} + 
  4\,{{a_{1}}}^4\,{{a_{1}^{*}}}^6\,{f^{*}}\,
   {{g'}}^2\,{{{g'}^{*}}}^2\,{\Sigma_{-}} + 
  4\,{{a_{1}}}^6\,{{a_{1}^{*}}}^4\,{f^{*}}\,
   {g'}\,{{{g'}^{*}}}^3\,{\Sigma_{-}} + 
  4\,{{a_{1}}}^8\,{{a_{1}^{*}}}^2\,{f^{*}}\,
   {{{g'}^{*}}}^4\,{\Sigma_{-}} + 
  6\,{{a_{1}}}^4\,{{a_{1}^{*}}}^6\,d\,f\,{{f^{*}}}^2\,
   {{\Sigma_{-}}}^2 + 4\,{{a_{1}}}^4\,{{a_{1}^{*}}}^6\,d\,
   {{f^{*}}}^3\,{{\Sigma_{-}}}^2 + 
  36\,{{a_{1}}}^2\,{{a_{1}^{*}}}^8\,{d^{*}}\,f\,
   {f^{*}}\,{g'}\,{{\Sigma_{-}}}^2 + 
  6\,{{a_{1}}}^2\,{{a_{1}^{*}}}^8\,{d^{*}}\,
   {{f^{*}}}^2\,{g'}\,{{\Sigma_{-}}}^2 + 
  6\,{{a_{1}}}^2\,{{a_{1}^{*}}}^8\,d\,{f^{*}}\,
   {{g'}}^2\,{{\Sigma_{-}}}^2 + 
  16\,{{a_{1}^{*}}}^{10}\,{d^{*}}\,{{g'}}^3\,
   {{\Sigma_{-}}}^2 + 24\,{{a_{1}}}^4\,{{a_{1}^{*}}}^6\,
   {d^{*}}\,f\,{f^{*}}\,{{g'}^{*}}\,{{\Sigma_{-}}}^2 + 
  4\,{{a_{1}}}^4\,{{a_{1}^{*}}}^6\,{d^{*}}\,
   {{f^{*}}}^2\,{{g'}^{*}}\,{{\Sigma_{-}}}^2 + 
  8\,{{a_{1}}}^4\,{{a_{1}^{*}}}^6\,d\,{f^{*}}\,
   {g'}\,{{g'}^{*}}\,{{\Sigma_{-}}}^2 + 
  12\,{{a_{1}}}^2\,{{a_{1}^{*}}}^8\,{d^{*}}\,
   {{g'}}^2\,{{g'}^{*}}\,{{\Sigma_{-}}}^2 + 
  6\,{{a_{1}}}^6\,{{a_{1}^{*}}}^4\,d\,{f^{*}}\,
   {{{g'}^{*}}}^2\,{{\Sigma_{-}}}^2 + 
  8\,{{a_{1}}}^4\,{{a_{1}^{*}}}^6\,{d^{*}}\,
   {g'}\,{{{g'}^{*}}}^2\,{{\Sigma_{-}}}^2 + 
  4\,{{a_{1}}}^6\,{{a_{1}^{*}}}^4\,{d^{*}}\,
   {{{g'}^{*}}}^3\,{{\Sigma_{-}}}^2 + 
  8\,{{a_{1}}}^2\,{{a_{1}^{*}}}^8\,{{d^{*}}}^2\,f\,
   {{\Sigma_{-}}}^3 + {{a_{1}}}^4\,{{a_{1}^{*}}}^6\,d^2\,
   {f^{*}}\,{{\Sigma_{-}}}^3 + 
  6\,{{a_{1}}}^2\,{{a_{1}^{*}}}^8\,{{d^{*}}}^2\,
   {f^{*}}\,{{\Sigma_{-}}}^3 + 
  6\,{{a_{1}}}^2\,{{a_{1}^{*}}}^8\,d\,{d^{*}}\,
   {g'}\,{{\Sigma_{-}}}^3 + 
  4\,{{a_{1}}}^4\,{{a_{1}^{*}}}^6\,d\,{d^{*}}\,
   {{g'}^{*}}\,{{\Sigma_{-}}}^3 + 
  8\,{{a_{1}}}^4\,{{a_{1}^{*}}}^6\,f^2\,{{f^{*}}}^2\,
   {g'}\,{\Sigma_{+}} + 
  4\,{{a_{1}}}^2\,{{a_{1}^{*}}}^8\,f\,{f^{*}}\,
   {{g'}}^3\,{\Sigma_{+}} + 
  12\,{{a_{1}}}^6\,{{a_{1}^{*}}}^4\,f^2\,{{f^{*}}}^2\,
   {{g'}^{*}}\,{\Sigma_{+}} + 
  8\,{{a_{1}}}^4\,{{a_{1}^{*}}}^6\,f\,{f^{*}}\,
   {{g'}}^2\,{{g'}^{*}}\,{\Sigma_{+}} + 
  12\,{{a_{1}}}^6\,{{a_{1}^{*}}}^4\,f\,{f^{*}}\,
   {g'}\,{{{g'}^{*}}}^2\,{\Sigma_{+}} + 
  16\,{{a_{1}}}^8\,{{a_{1}^{*}}}^2\,f\,{f^{*}}\,
   {{{g'}^{*}}}^3\,{\Sigma_{+}} + 
  4\,{{a_{1}}}^{10}\,{{{g'}^{*}}}^5\,{\Sigma_{+}} + 
  24\,{{a_{1}}}^4\,{{a_{1}^{*}}}^6\,{d^{*}}\,f^2\,
   {f^{*}}\,{\Sigma_{-}}\,{\Sigma_{+}} + 
  8\,{{a_{1}}}^4\,{{a_{1}^{*}}}^6\,{d^{*}}\,f\,
   {{f^{*}}}^2\,{\Sigma_{-}}\,{\Sigma_{+}} + 
  16\,{{a_{1}}}^4\,{{a_{1}^{*}}}^6\,d\,f\,{f^{*}}\,
   {g'}\,{\Sigma_{-}}\,{\Sigma_{+}} + 
  16\,{{a_{1}}}^4\,{{a_{1}^{*}}}^6\,d\,{{f^{*}}}^2\,
   {g'}\,{\Sigma_{-}}\,{\Sigma_{+}} + 
  24\,{{a_{1}}}^2\,{{a_{1}^{*}}}^8\,{d^{*}}\,f\,
   {{g'}}^2\,{\Sigma_{-}}\,{\Sigma_{+}} + 
  6\,{{a_{1}}}^2\,{{a_{1}^{*}}}^8\,{d^{*}}\,
   {f^{*}}\,{{g'}}^2\,{\Sigma_{-}}\,{\Sigma_{+}} + 
  2\,{{a_{1}}}^2\,{{a_{1}^{*}}}^8\,d\,{{g'}}^3\,
   {\Sigma_{-}}\,{\Sigma_{+}} + 
  24\,{{a_{1}}}^6\,{{a_{1}^{*}}}^4\,d\,f\,{f^{*}}\,
   {{g'}^{*}}\,{\Sigma_{-}}\,{\Sigma_{+}} + 
  24\,{{a_{1}}}^6\,{{a_{1}^{*}}}^4\,d\,{{f^{*}}}^2\,
   {{g'}^{*}}\,{\Sigma_{-}}\,{\Sigma_{+}} + 
  32\,{{a_{1}}}^4\,{{a_{1}^{*}}}^6\,{d^{*}}\,f\,
   {g'}\,{{g'}^{*}}\,{\Sigma_{-}}\,{\Sigma_{+}} + 
  8\,{{a_{1}}}^4\,{{a_{1}^{*}}}^6\,{d^{*}}\,
   {f^{*}}\,{g'}\,{{g'}^{*}}\,{\Sigma_{-}}\,
   {\Sigma_{+}} + 4\,{{a_{1}}}^4\,{{a_{1}^{*}}}^6\,d\,
   {{g'}}^2\,{{g'}^{*}}\,{\Sigma_{-}}\,{\Sigma_{+}} + 
  24\,{{a_{1}}}^6\,{{a_{1}^{*}}}^4\,{d^{*}}\,f\,
   {{{g'}^{*}}}^2\,{\Sigma_{-}}\,{\Sigma_{+}} + 
  6\,{{a_{1}}}^6\,{{a_{1}^{*}}}^4\,{d^{*}}\,
   {f^{*}}\,{{{g'}^{*}}}^2\,{\Sigma_{-}}\,{\Sigma_{+}} + 
  6\,{{a_{1}}}^6\,{{a_{1}^{*}}}^4\,d\,{g'}\,
   {{{g'}^{*}}}^2\,{\Sigma_{-}}\,{\Sigma_{+}} + 
  8\,{{a_{1}}}^8\,{{a_{1}^{*}}}^2\,d\,{{{g'}^{*}}}^3\,
   {\Sigma_{-}}\,{\Sigma_{+}} + 
  12\,{{a_{1}}}^4\,{{a_{1}^{*}}}^6\,d\,{d^{*}}\,f\,
   {{\Sigma_{-}}}^2\,{\Sigma_{+}} + 
  28\,{{a_{1}}}^4\,{{a_{1}^{*}}}^6\,d\,{d^{*}}\,
   {f^{*}}\,{{\Sigma_{-}}}^2\,{\Sigma_{+}} + 
  2\,{{a_{1}}}^4\,{{a_{1}^{*}}}^6\,d^2\,{g'}\,
   {{\Sigma_{-}}}^2\,{\Sigma_{+}} + 
  12\,{{a_{1}}}^2\,{{a_{1}^{*}}}^8\,{{d^{*}}}^2\,
   {g'}\,{{\Sigma_{-}}}^2\,{\Sigma_{+}} + 
  3\,{{a_{1}}}^6\,{{a_{1}^{*}}}^4\,d^2\,{{g'}^{*}}\,
   {{\Sigma_{-}}}^2\,{\Sigma_{+}} + 
  8\,{{a_{1}}}^4\,{{a_{1}^{*}}}^6\,{{d^{*}}}^2\,
   {{g'}^{*}}\,{{\Sigma_{-}}}^2\,{\Sigma_{+}} + 
  6\,{{a_{1}}}^6\,{{a_{1}^{*}}}^4\,d\,f^2\,{f^{*}}\,
   {{\Sigma_{+}}}^2 + 12\,{{a_{1}}}^6\,{{a_{1}^{*}}}^4\,d\,
   f\,{{f^{*}}}^2\,{{\Sigma_{+}}}^2 + 
  8\,{{a_{1}}}^4\,{{a_{1}^{*}}}^6\,{d^{*}}\,f^2\,
   {g'}\,{{\Sigma_{+}}}^2 + 
  4\,{{a_{1}}}^4\,{{a_{1}^{*}}}^6\,{d^{*}}\,f\,
   {f^{*}}\,{g'}\,{{\Sigma_{+}}}^2 + 
  2\,{{a_{1}}}^4\,{{a_{1}^{*}}}^6\,d\,f\,{{g'}}^2\,
   {{\Sigma_{+}}}^2 + 4\,{{a_{1}}}^4\,{{a_{1}^{*}}}^6\,d\,
   {f^{*}}\,{{g'}}^2\,{{\Sigma_{+}}}^2 + 
  12\,{{a_{1}}}^6\,{{a_{1}^{*}}}^4\,{d^{*}}\,f^2\,
   {{g'}^{*}}\,{{\Sigma_{+}}}^2 + 
  6\,{{a_{1}}}^6\,{{a_{1}^{*}}}^4\,{d^{*}}\,f\,
   {f^{*}}\,{{g'}^{*}}\,{{\Sigma_{+}}}^2 + 
  6\,{{a_{1}}}^6\,{{a_{1}^{*}}}^4\,d\,f\,{g'}\,
   {{g'}^{*}}\,{{\Sigma_{+}}}^2 + 
  12\,{{a_{1}}}^6\,{{a_{1}^{*}}}^4\,d\,{f^{*}}\,
   {g'}\,{{g'}^{*}}\,{{\Sigma_{+}}}^2 + 
  12\,{{a_{1}}}^8\,{{a_{1}^{*}}}^2\,d\,f\,{{{g'}^{*}}}^2\,
   {{\Sigma_{+}}}^2 + 24\,{{a_{1}}}^8\,{{a_{1}^{*}}}^2\,d\,
   {f^{*}}\,{{{g'}^{*}}}^2\,{{\Sigma_{+}}}^2 + 
  3\,{{a_{1}}}^6\,{{a_{1}^{*}}}^4\,d^2\,f\,{\Sigma_{-}}\,
   {{\Sigma_{+}}}^2 + 8\,{{a_{1}}}^4\,{{a_{1}^{*}}}^6\,
   {{d^{*}}}^2\,f\,{\Sigma_{-}}\,{{\Sigma_{+}}}^2 + 
  12\,{{a_{1}}}^6\,{{a_{1}^{*}}}^4\,d^2\,{f^{*}}\,
   {\Sigma_{-}}\,{{\Sigma_{+}}}^2 + 
  {{a_{1}}}^4\,{{a_{1}^{*}}}^6\,{{d^{*}}}^2\,
   {f^{*}}\,{\Sigma_{-}}\,{{\Sigma_{+}}}^2 + 
  18\,{{a_{1}}}^4\,{{a_{1}^{*}}}^6\,d\,{d^{*}}\,
   {g'}\,{\Sigma_{-}}\,{{\Sigma_{+}}}^2 + 
  27\,{{a_{1}}}^6\,{{a_{1}^{*}}}^4\,d\,{d^{*}}\,
   {{g'}^{*}}\,{\Sigma_{-}}\,{{\Sigma_{+}}}^2 + 
  9\,{{a_{1}}}^6\,{{a_{1}^{*}}}^4\,d\,{d^{*}}\,f\,
   {{\Sigma_{+}}}^3 + 2\,{{a_{1}}}^6\,{{a_{1}^{*}}}^4\,d\,
   {d^{*}}\,{f^{*}}\,{{\Sigma_{+}}}^3 + 
  2\,{{a_{1}}}^6\,{{a_{1}^{*}}}^4\,d^2\,{g'}\,
   {{\Sigma_{+}}}^3 + 8\,{{a_{1}}}^8\,{{a_{1}^{*}}}^2\,d^2\,
   {{g'}^{*}}\,{{\Sigma_{+}}}^3$.

\baselineskip=\normalbaselineskip\normalsize

\vspace{5mm}
These polynomials were found using a system of analytical computations
{\it Mathematica} \cite{Mathematica}.
]

\noindent
separation of $\Phi$ branches on the classical level
provides an alternative approach to this problem. Particularly,
the spectrum of $P^{2}$ in the vicinity of straight-line string
can be studied using the expansion (\ref{a2cl}). In this paper we will
consider this problem for a particular case of spin singlet $S=0$.

\vsp\subsection{Solutions in the vicinity of straight-line string}
Let's fix $s=2$ and put $S_{3}=S_{\pm}=0$ in the formulae above.
Quantum analog of (\ref{a2cl}) is constructed as follows:

\begin{eqnarray}
&&a_{2}=\sum\limits_{n\geq1}:P_{n}a_{1}^{2}:D^{-2n},\label{a2q}
\end{eqnarray}
where $D=a_{1}^{+}a_{1}+1/2=(a_{1}^{+}a_{1}+a_{1}a_{1}^{+})/2$.

Polynomials $P_{n}$ are defined by expressions of Appendix~1,
where we substitute the definitions (\ref{defelem}) of elementary operators
and fix the ordering, shown in Table~1.
This ordering puts $L_{0}^{(2)}$-lowering elementary operators
to the right from $L_{0}^{(2)}$-raising ones, thus providing better
convergence properties for the expansion (\ref{a2q}). Particularly,
the matrix elements of $a_{2}$ between the states with finite 
$L_{0}^{(2)}$ are given by finite sums, and large $n$ terms
of (\ref{a2q}) contribute only to the matrix elements with large 
$L_{0}^{(2)}$, due to the following lemma.

\vspace{2mm}
\noindent{\bf Lemma~7:}\\
$\bra{L_{0}^{(2)}=N_{1}} :P_{n}: \ket{L_{0}^{(2)}=N_{2}}=0$,
if $4(n-1)>N_{1}+N_{2}$.

\vspace{2mm}\noindent
Square-mass operator is given by
\begin{eqnarray}
&&{P^{2}\over2\pi}=L_{0}^{(2)}+2a_{2}^{+}a_{2}.\label{P2q}
\end{eqnarray}
One can also include here classically vanishing constant term:
$P^{2}\to P^{2}+c$. In this definition (at $c>0$) the operator $P^{2}$
is Hermitian and positively defined\footnote{
Note that other possible definition 
$P^{2}/2\pi=L_{0}^{(2)}+2:a_{2}^{+}a_{2}:$ does not have this property.}.

The introduced ordering possesses the other feature 
convenient for computations: the matrix elements of 
the normally ordered operators restricted to
finite-dimensional subspaces, behave regularly when
the dimension of subspaces increases. Namely, 
the matrix in larger space includes the matrix in
smaller space as an exact submatrix. Without normal ordering
this simple behavior would be violated.

\vspace{2mm}
\noindent{\bf Lemma~8:} let 
$op_{i}(N)=\bra{L_{0}^{(2)}\leq N}op_{i}\ket{L_{0}^{(2)}\leq N}$
be restrictions of elementary operators $op_{i}$ to finite-dimensional
space $L_{0}^{(2)}\leq N$. Let $:P(op_{i}):$
be $L_{0}^{(2)}$-normally ordered polynomial of $op_{i}$.
Then (i) at $N_{1}<N_{2}$ the matrix $:P(op_{i}(N_{1})):$ is
a submatrix of $:P(op_{i}(N_{2})):$ and (ii) $:P(op_{i}(N)):=
\bra{L_{0}^{(2)}\leq N}:P(op_{i}):\ket{L_{0}^{(2)}\leq N}$.

\begin{figure}\label{spec}
\begin{center}
~\epsfxsize=8cm\epsfysize=4cm\epsffile{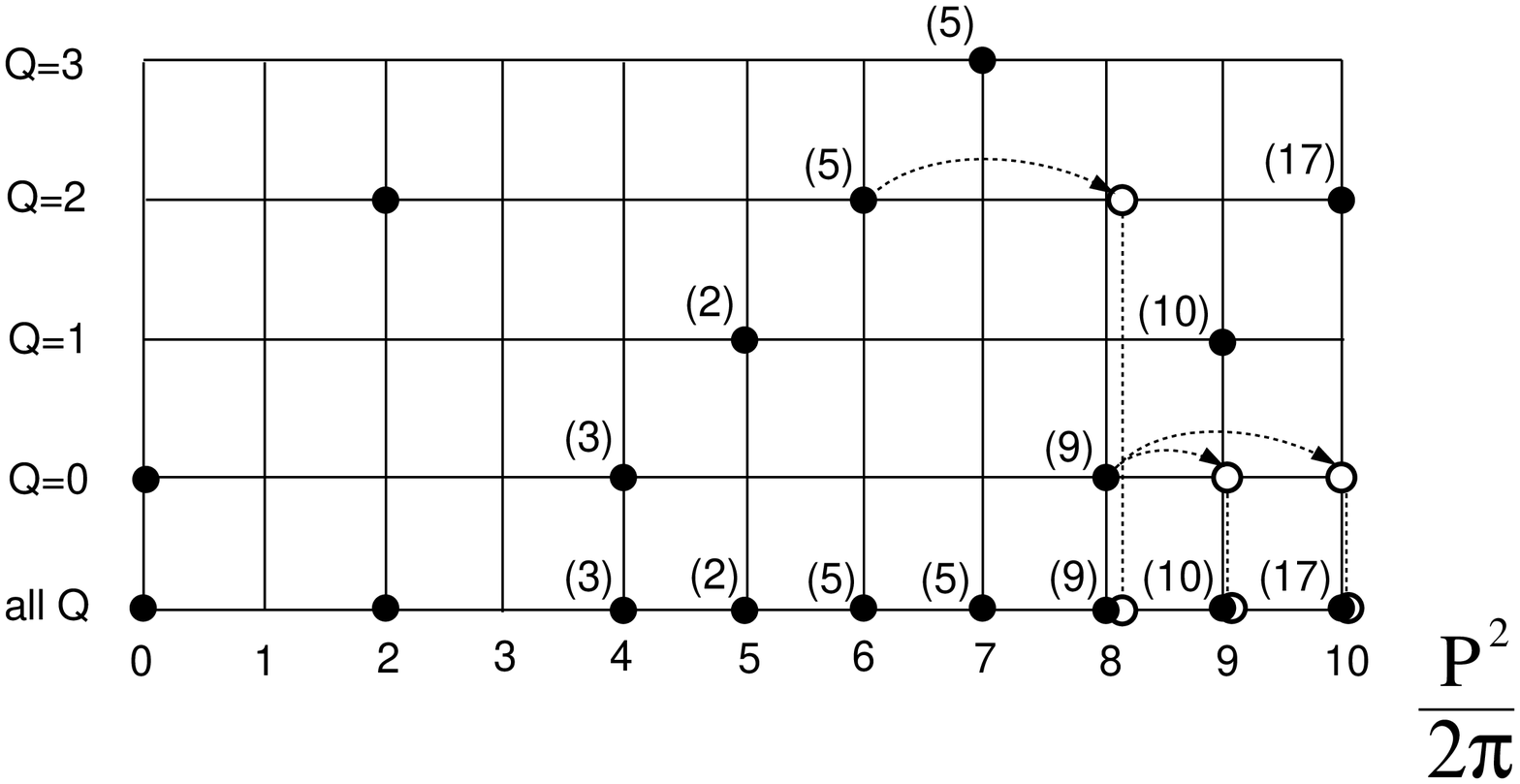}

\fignum Spectrum $(P^{2}/2\pi,S=0)$.
\end{center}
\end{figure}

\vspace{2mm}
Practically, we compute the matrix elements of elementary operators 
in $L_{0}^{(2)}\leq N$ subspace.
The dimension of this subspace is rapidly increasing
with $N$, e.g. $dim=106587$ for $N=25$,
so that the elementary operators are represented 
as matrices of very large size ($dim\times dim$),
see Appendix~3.
The matrices have noticeable block structure,
corresponding to their $(A_{3}^{(2)},L_{0}^{(2)})$-charge properties.
As a result, non-zero matrix elements of $a_{2}$, necessary
for computation of $P^{2}$ at $S=0$, are located in the blocks 
$\bra{A_{3}=-1,Q+2}a_{2}\ket{A_{3}=0,Q}$, while $P^{2}$
itself is located in $\bra{A_{3}=0,Q}P^{2}\ket{A_{3}=0,Q}$.
In addition to these properties we use the fact that elementary matrices 
inside $(A_{3}^{(2)},L_{0}^{(2)})$-blocks are very sparse
(at large $N$ their non-zero content occupies less than 1\% of the blocks),
and implement special algorithms for sparse block matrix computations,
described in more details in Appendix~3. Finally, we determine the
spectrum of $P^{2}/2\pi$ up to the value $N=28$ ($dim=260256$) and 
the number of terms in expansion (\ref{a2q}) up to $n=6$.
The initial part of $P^{2}$ spectrum is rapidly stabilized when
$n$ is increasing. Particularly, the region $P^{2}/2\pi\leq10$
is defined by $n\leq3$ terms, while $n>3$ corrections
influence the higher $P^{2}$ values only. Figure \ref{spec}
shows the resulting spectrum of $P^{2}$, for separate $Q$-values 
and superimposed. 

Table~2: eigenvectors $(P^{2}/2\pi\leq5,S=0)$.

$$\begin{array}{|c|c|}\hline
P^{2}/2\pi=L_{0}^{(2)}&\ket{\{n_{k}\}}\\\hline
0&\ket{0}\\
2&\ket{1_{1}1_{-1}}\\
4&\ket{2_{1}2_{-1}},\ket{1_{3}1_{-1}},\ket{1_{1}1_{-3}}\\
5&\ket{1_{4}1_{-1}},\ket{1_{1}1_{-4}}\\\hline
\end{array}
$$

The spectrum contains degenerate eigenvalues,
whose multiplicities are shown on the figure by numbers in brackets
(non-degenerate eigenvalues have no these numbers).
Black points display the states annulated by operator $a_{2}$,
which are simultaneously the eigenvectors of $L_{0}^{(2)}$
and therefore have integer $P^{2}/2\pi$. The beginning of the spectrum
consists only of such states, as the following lemma shows.

\vspace{2mm}
\noindent{\bf Lemma~9:} $a_{2}$ annulates $\ket{L_{0}^{(2)}\leq5}$.

\vspace{2mm}
At $P^{2}/2\pi\geq6$ the eigenvalues of $L_{0}^{(2)}$ start to split.
Particularly, one state with $L_{0}^{(2)}=6$ goes to $P^{2}/2\pi=8.16792$,
two states with $L_{0}^{(2)}=8$ become $P^{2}/2\pi=9.0107,10.018$ etc.
Such states (displayed by white points on \fref{spec}) 
correspond to the exited $a_{2}$-modes. The values of $P^{2}/2\pi$
for these states are not integer (however, some of them
are numerically close to integer). Many eigenvectors of 
$L_{0}^{(2)}\geq6$ are also annulated by $a_{2}$, their $P^{2}/2\pi$
are degenerate and integer. This degeneracy can be removed
by further model corrections, such as spin-orbital interaction 
discussed in \cite{slstring}.


Subspaces with different $Q$ in quantum theory play a role
of factor-spaces with respect to $D_{4}$-symmetry,
which is satisfied exactly for $Q=0$: $D_{4}\Psi=\Psi$, 
and in projective sense for $Q\neq0$: $D_{4}\Psi\sim\Psi$.
Quantum Hamiltonians of interaction, possessing $D_{4}$-symmetry,
keep these subspaces invariant. Hamiltonians, violating this
symmetry, create transitions between different $Q$-sectors.
If the violation of $D_{4}$-symmetry is small, the particles
on the first daughter Regge trajectories will be quasi-stable.


\vsp\paragraph*{Summary of quantum mechanics:}
representation of canonical operators (\ref{canon}) is straightforward.
Quantum Hamiltonian can be defined either as polynomial function 
(Appendix~1) or as expansion (\ref{a2q}) in the vicinity
of straight-line string, both properly ordered.
In the first definition a non-linear spectral problem appears,
with associated separation task for the roots of characteristic 
equation. The second definition leads to the ordinary linear 
spectral problem, which allows, particularly, to determine 
the mass spectrum for the states of spin singlet $S=0$.

\vspace{3mm}
A paper devoted to systematic study of Gribov's copies
in this mechanics and determination of mass spectrum 
outside of the spin singlet will be published.


\def\figpage{
\vsp\paragraph*{Appendix 1:} the polynomials, entering
to the definition of Groebner's basis (\ref{groeb}) of the system 
(\ref{maineqs1}):

\vspace{5mm}
\baselineskip=0.4\normalbaselineskip\footnotesize

$H=
- {f^{*}}\,g\,{g^{*}}^2\,k^2 - f\,g^2\,{g^{*}}\,{k^{*}}^2 
- d\,{f^{*}}\,g\,k\,{k^{*}}^2 - {d^{*}}\,f\,{g^{*}}\,k^2\,{k^{*}} 
+ {d^{*}}\,{g^{*}}^2\,k^3 + d\,g^2\,{k^{*}}^3 
+ {d^{*}}\,f^2\,{f^{*}}\,{g^{*}}\,k\,n_{s} 
+ d\,f\,{f^{*}}^2\,g\,{k^{*}}\,n_{s}
+ 2\,{d^{*}}\,f^2\,g\,{g^{*}}\,{k^{*}}\,n_{s} 
+ 2\,d\,{f^{*}}^2\,g\,{g^{*}}\,k\,n_{s} 
- 3\,{d^{*}}\,f\,g\,{g^{*}}^2\,k\,n_{s} 
- 3\,d\,{f^{*}}\,g^2\,{g^{*}}\,{k^{*}}\,n_{s} 
- 2\,d\,{d^{*}}\,{f^{*}}\,{g^{*}}\,k^2\,n_{s} 
- 2\,d\,{d^{*}}\,f\,g\,{k^{*}}^2\,n_{s} 
- d^2\,{f^{*}}^3\,g\,n_{s}^2 
- {d^{*}}^2\,f^3\,{g^{*}}\,n_{s}^2  
+ d^2\,{d^{*}}\,{f^{*}}^2\,k\,n_{s}^2 
+ d\,{d^{*}}^2\,f^2\,{k^{*}}\,n_{s}^2  
+ 3\,d\,{d^{*}}^2\,f\,{g^{*}}\,k\,n_{s}^2 
+ 3\,d^2\,{d^{*}}\,{f^{*}}\,g\,{k^{*}}\,n_{s}^2  
+ f\,{f^{*}}\,g\,{g^{*}}\,k\,{k^{*}} 
+ g^2\,{g^{*}}^2\,k\,{k^{*}} 
+ d\,{d^{*}}\,k^2\,{k^{*}}^2  
- g^3\,{g^{*}}^3\,n_{s} 
+ 2\,f\,{f^{*}}\,g^2\,{g^{*}}^2\,n_{s} 
- f^2\,{f^{*}}^2\,g\,{g^{*}}\,n_{s}  
- d\,{d^{*}}\,f\,{f^{*}}\,k\,{k^{*}}\,n_{s} 
+ d\,{d^{*}}\,g\,{g^{*}}\,k\,{k^{*}}\,n_{s}  
- d\,{d^{*}}\,f\,{f^{*}}\,g\,{g^{*}}\,n_{s}^2 
+ 3\,d\,{d^{*}}\,g^2\,{g^{*}}^2\,n_{s}^2 
- 2\,d^2\,{d^{*}}^2\,k\,{k^{*}}\,n_{s}^2  
- 3\,d^2\,{d^{*}}^2\,g\,{g^{*}}\,n_{s}^3 
- d^2\,{d^{*}}^2\,f\,{f^{*}}\,n_{s}^3  
+ d^3\,{d^{*}}^3\,n_{s}^4 
$

\vspace{2mm}
$F=
-{f^{*}}^3\,g^2\,{g^{*}}\,k\,n_{s}
+ 2\,{d^{*}}\,{f^{*}}^2\,g\,{g^{*}}\,k^2\,n_{s} 
- {d^{*}}^2\,{f^{*}}\,{g^{*}}\,k^3\,n_{s} 
+ {f^{*}}^2\,g^3\,{g^{*}}\,{k^{*}}\,n_{s} 
- {d^{*}}^2\,g\,{g^{*}}\,k^2\,{k^{*}}\,n_{s} 
- {d^{*}}\,g^3\,{g^{*}}\,{k^{*}}^2\,n_{s} 
+ d\,{f^{*}}^4\,g^2\,n_{s}^2 
- {d^{*}}\,{f^{*}}\,g^3\,{g^{*}}^2\,n_{s}^2 
- 2\,d\,{d^{*}}\,{f^{*}}^3\,g\,k\,n_{s}^2 
- 2\,{d^{*}}^2\,f\,{f^{*}}\,g\,{g^{*}}\,k\,n_{s}^2 
+ {d^{*}}^2\,g^2\,{g^{*}}^2\,k\,n_{s}^2 
+ d\,{d^{*}}^2\,{f^{*}}^2\,k^2\,n_{s}^2 
+ 2\,{d^{*}}^3\,f\,{g^{*}}\,k^2\,n_{s}^2 
- 3\,d\,{d^{*}}\,{f^{*}}^2\,g^2\,{k^{*}}\,n_{s}^2 
+ 2\,{d^{*}}^2\,f\,g^2\,{g^{*}}\,{k^{*}}\, n_{s}^2 
+ 4\,d\,{d^{*}}^2\,{f^{*}}\,g\,k\,{k^{*}}\,n_{s}^2 
- d\,{d^{*}}^3\,k^2\,{k^{*}}\,n_{s}^2 
+ d\,{d^{*}}^2\,g^2\,{k^{*}}^2\,n_{s}^2 
+ 2\,d\,{d^{*}}^2\,f\,{f^{*}}^2\,g\,n_{s}^3 
- {d^{*}}^3\,f^2\,g\,{g^{*}}\,n_{s}^3 
+ 2\,d\,{d^{*}}^2\,{f^{*}}\,g^2\,{g^{*}}\,n_{s}^3 
- 2\,d\,{d^{*}}^3\,f\,{f^{*}}\,k\,n_{s}^3 
- 2\,d\,{d^{*}}^3\,g\,{g^{*}}\,k\,n_{s}^3 
- 2\,d\,{d^{*}}^3\,f\,g\,{k^{*}}\,n_{s}^3 
+ d\,{d^{*}}^4\,f^2\,n_{s}^4 
- d^2\,{d^{*}}^3\,{f^{*}}\,g\,n_{s}^4 
+ d^2\,{d^{*}}^4\,k\,n_{s}^4$

\vspace{2mm}
$G={f^{*}}^3\,g^2\,k\,{k^{*}} 
- 2\,{d^{*}}\,{f^{*}}^2\,g\,k^2\,{k^{*}} 
+ {d^{*}}^2\,{f^{*}}\,k^3\,{k^{*}} 
- {f^{*}}^2\,g^3\,{k^{*}}^2 
+ {d^{*}}^2\,g\,k^2\,{k^{*}}^2 
+ {d^{*}}\,g^3\,{k^{*}}^3 
- f\,{f^{*}}^4\,g^2\,n_{s} 
+ {f^{*}}^3\,g^3\,{g^{*}}\,n_{s} 
+ 2\,{d^{*}}\,f\,{f^{*}}^3\,g\,k\,n_{s} 
- 3\,{d^{*}}\,{f^{*}}^2\,g^2\,{g^{*}}\,k\,n_{s} 
- {d^{*}}^2\,f\,{f^{*}}^2\,k^2\,n_{s} 
+ 3\,{d^{*}}^2\,{f^{*}}\,g\,{g^{*}}\,k^2\,n_{s} 
- {d^{*}}^3\,{g^{*}}\,k^3\,n_{s} 
+ 3\,{d^{*}}\,f\,{f^{*}}^2\,g^2\,{k^{*}}\,n_{s} 
- 2\,{d^{*}}^2\,f\,{f^{*}}\,g\,k\,{k^{*}}\,n_{s} 
- {d^{*}}^3\,f\,k^2\,{k^{*}}\,n_{s} 
- 3\,{d^{*}}^2\,f\,g^2\,{k^{*}}^2\,n_{s} 
- 2\,{d^{*}}^2\,f^2\,{f^{*}}^2\,g\,n_{s}^2 
+ 2\,{d^{*}}^3\,f^2\,{f^{*}}\,k\,n_{s}^2 
+ 3\,{d^{*}}^3\,f^2\,g\,{k^{*}}\,n_{s}^2 
- {d^{*}}^4\,f^3\,n_{s}^3$

\baselineskip=\normalbaselineskip\normalsize

\vspace{5mm}
The polynomials, entering to the expansion (\ref{a2cl})
of variable $a_{2}|_{S_{\pm}=0}$ in the vicinity of straight-line
string solution (\ref{1mode}):

\vspace{5mm}
\baselineskip=0.4\normalbaselineskip\footnotesize

$P_{1}=-\Sigma_{+},$

\vspace{2mm}
$P_{2}={{a_{1}^{*}}}^2\,{f^{*}}\,{\Sigma_{-}} + 
  {{a_{1}}}^2\,{{g'}^{*}}\,{\Sigma_{+}},$

\vspace{2mm}
$2P_{3}=
-2\,{{a_{1}^{*}}}^4\,{f^{*}}\,{g'}\,{\Sigma_{-}} - 
  2\,{{a_{1}}}^2\,{{a_{1}^{*}}}^2\,{f^{*}}\,
   {{g'}^{*}}\,{\Sigma_{-}} - 
  2\,{{a_{1}^{*}}}^4\,{d^{*}}\,{{\Sigma_{-}}}^2 - 
  2\,{{a_{1}}}^2\,{{a_{1}^{*}}}^2\,f\,{f^{*}}\,
   {\Sigma_{+}} - 2\,{{a_{1}}}^4\,{{{g'}^{*}}}^2\,
   {\Sigma_{+}} - {{a_{1}}}^2\,{{a_{1}^{*}}}^2\,d\,
   {\Sigma_{-}}\,{\Sigma_{+}}$,
 
\vspace{2mm}
$2P_{4}=
2\,{{a_{1}}}^2\,{{a_{1}^{*}}}^4\,f\,{{f^{*}}}^2\,
   {\Sigma_{-}} + 2\,{{a_{1}^{*}}}^6\,{f^{*}}\,
   {{g'}}^2\,{\Sigma_{-}} + 
  2\,{{a_{1}}}^2\,{{a_{1}^{*}}}^4\,{f^{*}}\,
   {g'}\,{{g'}^{*}}\,{\Sigma_{-}} + 
  2\,{{a_{1}}}^4\,{{a_{1}^{*}}}^2\,{f^{*}}\,
   {{{g'}^{*}}}^2\,{\Sigma_{-}} + 
  {{a_{1}}}^2\,{{a_{1}^{*}}}^4\,d\,{f^{*}}\,
   {{\Sigma_{-}}}^2 + 4\,{{a_{1}^{*}}}^6\,{d^{*}}\,
   {g'}\,{{\Sigma_{-}}}^2 + 
  2\,{{a_{1}}}^2\,{{a_{1}^{*}}}^4\,{d^{*}}\,
   {{g'}^{*}}\,{{\Sigma_{-}}}^2 + 
  2\,{{a_{1}}}^2\,{{a_{1}^{*}}}^4\,f\,{f^{*}}\,
   {g'}\,{\Sigma_{+}} + 
  4\,{{a_{1}}}^4\,{{a_{1}^{*}}}^2\,f\,{f^{*}}\,
   {{g'}^{*}}\,{\Sigma_{+}} + 
  2\,{{a_{1}}}^6\,{{{g'}^{*}}}^3\,{\Sigma_{+}} + 
  4\,{{a_{1}}}^2\,{{a_{1}^{*}}}^4\,{d^{*}}\,f\,
   {\Sigma_{-}}\,{\Sigma_{+}} + 
  {{a_{1}}}^2\,{{a_{1}^{*}}}^4\,{d^{*}}\,{f^{*}}\,
   {\Sigma_{-}}\,{\Sigma_{+}} + 
  {{a_{1}}}^2\,{{a_{1}^{*}}}^4\,d\,{g'}\,
   {\Sigma_{-}}\,{\Sigma_{+}} + 
  2\,{{a_{1}}}^4\,{{a_{1}^{*}}}^2\,d\,{{g'}^{*}}\,
   {\Sigma_{-}}\,{\Sigma_{+}} + 
  {{a_{1}}}^4\,{{a_{1}^{*}}}^2\,d\,f\,{{\Sigma_{+}}}^2 + 
  2\,{{a_{1}}}^4\,{{a_{1}^{*}}}^2\,d\,{f^{*}}\,
   {{\Sigma_{+}}}^2$,
 
\vspace{2mm}
$4P_{5}=
-8\,{{a_{1}}}^2\,{{a_{1}^{*}}}^6\,f\,{{f^{*}}}^2\,
   {g'}\,{\Sigma_{-}} - 
  4\,{{a_{1}^{*}}}^8\,{f^{*}}\,{{g'}}^3\,
   {\Sigma_{-}} - 8\,{{a_{1}}}^4\,{{a_{1}^{*}}}^4\,f\,
   {{f^{*}}}^2\,{{g'}^{*}}\,{\Sigma_{-}} - 
  4\,{{a_{1}}}^2\,{{a_{1}^{*}}}^6\,{f^{*}}\,
   {{g'}}^2\,{{g'}^{*}}\,{\Sigma_{-}} - 
  4\,{{a_{1}}}^4\,{{a_{1}^{*}}}^4\,{f^{*}}\,
   {g'}\,{{{g'}^{*}}}^2\,{\Sigma_{-}} - 
  4\,{{a_{1}}}^6\,{{a_{1}^{*}}}^2\,{f^{*}}\,
   {{{g'}^{*}}}^3\,{\Sigma_{-}} - 
  12\,{{a_{1}}}^2\,{{a_{1}^{*}}}^6\,{d^{*}}\,f\,
   {f^{*}}\,{{\Sigma_{-}}}^2 - 
  2\,{{a_{1}}}^2\,{{a_{1}^{*}}}^6\,{d^{*}}\,
   {{f^{*}}}^2\,{{\Sigma_{-}}}^2 - 
  4\,{{a_{1}}}^2\,{{a_{1}^{*}}}^6\,d\,{f^{*}}\,
   {g'}\,{{\Sigma_{-}}}^2 - 
  12\,{{a_{1}^{*}}}^8\,{d^{*}}\,{{g'}}^2\,
   {{\Sigma_{-}}}^2 - 4\,{{a_{1}}}^4\,{{a_{1}^{*}}}^4\,d\,
   {f^{*}}\,{{g'}^{*}}\,{{\Sigma_{-}}}^2 - 
  8\,{{a_{1}}}^2\,{{a_{1}^{*}}}^6\,{d^{*}}\,
   {g'}\,{{g'}^{*}}\,{{\Sigma_{-}}}^2 - 
  4\,{{a_{1}}}^4\,{{a_{1}^{*}}}^4\,{d^{*}}\,
   {{{g'}^{*}}}^2\,{{\Sigma_{-}}}^2 - 
  2\,{{a_{1}}}^2\,{{a_{1}^{*}}}^6\,d\,{d^{*}}\,
   {{\Sigma_{-}}}^3 - 4\,{{a_{1}}}^4\,{{a_{1}^{*}}}^4\,f^2\,
   {{f^{*}}}^2\,{\Sigma_{+}} - 
  4\,{{a_{1}}}^2\,{{a_{1}^{*}}}^6\,f\,{f^{*}}\,
   {{g'}}^2\,{\Sigma_{+}} - 
  8\,{{a_{1}}}^4\,{{a_{1}^{*}}}^4\,f\,{f^{*}}\,
   {g'}\,{{g'}^{*}}\,{\Sigma_{+}} - 
  12\,{{a_{1}}}^6\,{{a_{1}^{*}}}^2\,f\,{f^{*}}\,
   {{{g'}^{*}}}^2\,{\Sigma_{+}} - 
  4\,{{a_{1}}}^8\,{{{g'}^{*}}}^4\,{\Sigma_{+}} - 
  8\,{{a_{1}}}^4\,{{a_{1}^{*}}}^4\,d\,f\,{f^{*}}\,
   {\Sigma_{-}}\,{\Sigma_{+}} - 
  8\,{{a_{1}}}^4\,{{a_{1}^{*}}}^4\,d\,{{f^{*}}}^2\,
   {\Sigma_{-}}\,{\Sigma_{+}} - 
  16\,{{a_{1}}}^2\,{{a_{1}^{*}}}^6\,{d^{*}}\,f\,
   {g'}\,{\Sigma_{-}}\,{\Sigma_{+}} - 
  4\,{{a_{1}}}^2\,{{a_{1}^{*}}}^6\,{d^{*}}\,
   {f^{*}}\,{g'}\,{\Sigma_{-}}\,{\Sigma_{+}} - 
  2\,{{a_{1}}}^2\,{{a_{1}^{*}}}^6\,d\,{{g'}}^2\,
   {\Sigma_{-}}\,{\Sigma_{+}} - 
  16\,{{a_{1}}}^4\,{{a_{1}^{*}}}^4\,{d^{*}}\,f\,
   {{g'}^{*}}\,{\Sigma_{-}}\,{\Sigma_{+}} - 
  4\,{{a_{1}}}^4\,{{a_{1}^{*}}}^4\,{d^{*}}\,
   {f^{*}}\,{{g'}^{*}}\,{\Sigma_{-}}\,{\Sigma_{+}} - 
  4\,{{a_{1}}}^4\,{{a_{1}^{*}}}^4\,d\,{g'}\,
   {{g'}^{*}}\,{\Sigma_{-}}\,{\Sigma_{+}} - 
  6\,{{a_{1}}}^6\,{{a_{1}^{*}}}^2\,d\,{{{g'}^{*}}}^2\,
   {\Sigma_{-}}\,{\Sigma_{+}} - 
  {{a_{1}}}^4\,{{a_{1}^{*}}}^4\,d^2\,{{\Sigma_{-}}}^2\,
   {\Sigma_{+}} - 4\,{{a_{1}}}^2\,{{a_{1}^{*}}}^6\,
   {{d^{*}}}^2\,{{\Sigma_{-}}}^2\,{\Sigma_{+}} - 
  4\,{{a_{1}}}^4\,{{a_{1}^{*}}}^4\,{d^{*}}\,f^2\,
   {{\Sigma_{+}}}^2 - 2\,{{a_{1}}}^4\,{{a_{1}^{*}}}^4\,
   {d^{*}}\,f\,{f^{*}}\,{{\Sigma_{+}}}^2 - 
  2\,{{a_{1}}}^4\,{{a_{1}^{*}}}^4\,d\,f\,{g'}\,
   {{\Sigma_{+}}}^2 - 4\,{{a_{1}}}^4\,{{a_{1}^{*}}}^4\,d\,
   {f^{*}}\,{g'}\,{{\Sigma_{+}}}^2 - 
  6\,{{a_{1}}}^6\,{{a_{1}^{*}}}^2\,d\,f\,{{g'}^{*}}\,
   {{\Sigma_{+}}}^2 - 12\,{{a_{1}}}^6\,{{a_{1}^{*}}}^2\,d\,
   {f^{*}}\,{{g'}^{*}}\,{{\Sigma_{+}}}^2 - 
  9\,{{a_{1}}}^4\,{{a_{1}^{*}}}^4\,d\,{d^{*}}\,
   {\Sigma_{-}}\,{{\Sigma_{+}}}^2 - 
  2\,{{a_{1}}}^6\,{{a_{1}^{*}}}^2\,d^2\,{{\Sigma_{+}}}^3$,
 
\vspace{2mm}
$4P_{6}=
4\,{{a_{1}}}^4\,{{a_{1}^{*}}}^6\,f^2\,{{f^{*}}}^3\,
   {\Sigma_{-}} + 12\,{{a_{1}}}^2\,{{a_{1}^{*}}}^8\,f\,
   {{f^{*}}}^2\,{{g'}}^2\,{\Sigma_{-}} + 
  4\,{{a_{1}^{*}}}^{10}\,{f^{*}}\,{{g'}}^4\,
   {\Sigma_{-}} + 16\,{{a_{1}}}^4\,{{a_{1}^{*}}}^6\,f\,
   {{f^{*}}}^2\,{g'}\,{{g'}^{*}}\,{\Sigma_{-}} + 
  4\,{{a_{1}}}^2\,{{a_{1}^{*}}}^8\,{f^{*}}\,
   {{g'}}^3\,{{g'}^{*}}\,{\Sigma_{-}} + 
  12\,{{a_{1}}}^6\,{{a_{1}^{*}}}^4\,f\,{{f^{*}}}^2\,
   {{{g'}^{*}}}^2\,{\Sigma_{-}} + 
  4\,{{a_{1}}}^4\,{{a_{1}^{*}}}^6\,{f^{*}}\,
   {{g'}}^2\,{{{g'}^{*}}}^2\,{\Sigma_{-}} + 
  4\,{{a_{1}}}^6\,{{a_{1}^{*}}}^4\,{f^{*}}\,
   {g'}\,{{{g'}^{*}}}^3\,{\Sigma_{-}} + 
  4\,{{a_{1}}}^8\,{{a_{1}^{*}}}^2\,{f^{*}}\,
   {{{g'}^{*}}}^4\,{\Sigma_{-}} + 
  6\,{{a_{1}}}^4\,{{a_{1}^{*}}}^6\,d\,f\,{{f^{*}}}^2\,
   {{\Sigma_{-}}}^2 + 4\,{{a_{1}}}^4\,{{a_{1}^{*}}}^6\,d\,
   {{f^{*}}}^3\,{{\Sigma_{-}}}^2 + 
  36\,{{a_{1}}}^2\,{{a_{1}^{*}}}^8\,{d^{*}}\,f\,
   {f^{*}}\,{g'}\,{{\Sigma_{-}}}^2 + 
  6\,{{a_{1}}}^2\,{{a_{1}^{*}}}^8\,{d^{*}}\,
   {{f^{*}}}^2\,{g'}\,{{\Sigma_{-}}}^2 + 
  6\,{{a_{1}}}^2\,{{a_{1}^{*}}}^8\,d\,{f^{*}}\,
   {{g'}}^2\,{{\Sigma_{-}}}^2 + 
  16\,{{a_{1}^{*}}}^{10}\,{d^{*}}\,{{g'}}^3\,
   {{\Sigma_{-}}}^2 + 24\,{{a_{1}}}^4\,{{a_{1}^{*}}}^6\,
   {d^{*}}\,f\,{f^{*}}\,{{g'}^{*}}\,{{\Sigma_{-}}}^2 + 
  4\,{{a_{1}}}^4\,{{a_{1}^{*}}}^6\,{d^{*}}\,
   {{f^{*}}}^2\,{{g'}^{*}}\,{{\Sigma_{-}}}^2 + 
  8\,{{a_{1}}}^4\,{{a_{1}^{*}}}^6\,d\,{f^{*}}\,
   {g'}\,{{g'}^{*}}\,{{\Sigma_{-}}}^2 + 
  12\,{{a_{1}}}^2\,{{a_{1}^{*}}}^8\,{d^{*}}\,
   {{g'}}^2\,{{g'}^{*}}\,{{\Sigma_{-}}}^2 + 
  6\,{{a_{1}}}^6\,{{a_{1}^{*}}}^4\,d\,{f^{*}}\,
   {{{g'}^{*}}}^2\,{{\Sigma_{-}}}^2 + 
  8\,{{a_{1}}}^4\,{{a_{1}^{*}}}^6\,{d^{*}}\,
   {g'}\,{{{g'}^{*}}}^2\,{{\Sigma_{-}}}^2 + 
  4\,{{a_{1}}}^6\,{{a_{1}^{*}}}^4\,{d^{*}}\,
   {{{g'}^{*}}}^3\,{{\Sigma_{-}}}^2 + 
  8\,{{a_{1}}}^2\,{{a_{1}^{*}}}^8\,{{d^{*}}}^2\,f\,
   {{\Sigma_{-}}}^3 + {{a_{1}}}^4\,{{a_{1}^{*}}}^6\,d^2\,
   {f^{*}}\,{{\Sigma_{-}}}^3 + 
  6\,{{a_{1}}}^2\,{{a_{1}^{*}}}^8\,{{d^{*}}}^2\,
   {f^{*}}\,{{\Sigma_{-}}}^3 + 
  6\,{{a_{1}}}^2\,{{a_{1}^{*}}}^8\,d\,{d^{*}}\,
   {g'}\,{{\Sigma_{-}}}^3 + 
  4\,{{a_{1}}}^4\,{{a_{1}^{*}}}^6\,d\,{d^{*}}\,
   {{g'}^{*}}\,{{\Sigma_{-}}}^3 + 
  8\,{{a_{1}}}^4\,{{a_{1}^{*}}}^6\,f^2\,{{f^{*}}}^2\,
   {g'}\,{\Sigma_{+}} + 
  4\,{{a_{1}}}^2\,{{a_{1}^{*}}}^8\,f\,{f^{*}}\,
   {{g'}}^3\,{\Sigma_{+}} + 
  12\,{{a_{1}}}^6\,{{a_{1}^{*}}}^4\,f^2\,{{f^{*}}}^2\,
   {{g'}^{*}}\,{\Sigma_{+}} + 
  8\,{{a_{1}}}^4\,{{a_{1}^{*}}}^6\,f\,{f^{*}}\,
   {{g'}}^2\,{{g'}^{*}}\,{\Sigma_{+}} + 
  12\,{{a_{1}}}^6\,{{a_{1}^{*}}}^4\,f\,{f^{*}}\,
   {g'}\,{{{g'}^{*}}}^2\,{\Sigma_{+}} + 
  16\,{{a_{1}}}^8\,{{a_{1}^{*}}}^2\,f\,{f^{*}}\,
   {{{g'}^{*}}}^3\,{\Sigma_{+}} + 
  4\,{{a_{1}}}^{10}\,{{{g'}^{*}}}^5\,{\Sigma_{+}} + 
  24\,{{a_{1}}}^4\,{{a_{1}^{*}}}^6\,{d^{*}}\,f^2\,
   {f^{*}}\,{\Sigma_{-}}\,{\Sigma_{+}} + 
  8\,{{a_{1}}}^4\,{{a_{1}^{*}}}^6\,{d^{*}}\,f\,
   {{f^{*}}}^2\,{\Sigma_{-}}\,{\Sigma_{+}} + 
  16\,{{a_{1}}}^4\,{{a_{1}^{*}}}^6\,d\,f\,{f^{*}}\,
   {g'}\,{\Sigma_{-}}\,{\Sigma_{+}} + 
  16\,{{a_{1}}}^4\,{{a_{1}^{*}}}^6\,d\,{{f^{*}}}^2\,
   {g'}\,{\Sigma_{-}}\,{\Sigma_{+}} + 
  24\,{{a_{1}}}^2\,{{a_{1}^{*}}}^8\,{d^{*}}\,f\,
   {{g'}}^2\,{\Sigma_{-}}\,{\Sigma_{+}} + 
  6\,{{a_{1}}}^2\,{{a_{1}^{*}}}^8\,{d^{*}}\,
   {f^{*}}\,{{g'}}^2\,{\Sigma_{-}}\,{\Sigma_{+}} + 
  2\,{{a_{1}}}^2\,{{a_{1}^{*}}}^8\,d\,{{g'}}^3\,
   {\Sigma_{-}}\,{\Sigma_{+}} + 
  24\,{{a_{1}}}^6\,{{a_{1}^{*}}}^4\,d\,f\,{f^{*}}\,
   {{g'}^{*}}\,{\Sigma_{-}}\,{\Sigma_{+}} + 
  24\,{{a_{1}}}^6\,{{a_{1}^{*}}}^4\,d\,{{f^{*}}}^2\,
   {{g'}^{*}}\,{\Sigma_{-}}\,{\Sigma_{+}} + 
  32\,{{a_{1}}}^4\,{{a_{1}^{*}}}^6\,{d^{*}}\,f\,
   {g'}\,{{g'}^{*}}\,{\Sigma_{-}}\,{\Sigma_{+}} + 
  8\,{{a_{1}}}^4\,{{a_{1}^{*}}}^6\,{d^{*}}\,
   {f^{*}}\,{g'}\,{{g'}^{*}}\,{\Sigma_{-}}\,
   {\Sigma_{+}} + 4\,{{a_{1}}}^4\,{{a_{1}^{*}}}^6\,d\,
   {{g'}}^2\,{{g'}^{*}}\,{\Sigma_{-}}\,{\Sigma_{+}} + 
  24\,{{a_{1}}}^6\,{{a_{1}^{*}}}^4\,{d^{*}}\,f\,
   {{{g'}^{*}}}^2\,{\Sigma_{-}}\,{\Sigma_{+}} + 
  6\,{{a_{1}}}^6\,{{a_{1}^{*}}}^4\,{d^{*}}\,
   {f^{*}}\,{{{g'}^{*}}}^2\,{\Sigma_{-}}\,{\Sigma_{+}} + 
  6\,{{a_{1}}}^6\,{{a_{1}^{*}}}^4\,d\,{g'}\,
   {{{g'}^{*}}}^2\,{\Sigma_{-}}\,{\Sigma_{+}} + 
  8\,{{a_{1}}}^8\,{{a_{1}^{*}}}^2\,d\,{{{g'}^{*}}}^3\,
   {\Sigma_{-}}\,{\Sigma_{+}} + 
  12\,{{a_{1}}}^4\,{{a_{1}^{*}}}^6\,d\,{d^{*}}\,f\,
   {{\Sigma_{-}}}^2\,{\Sigma_{+}} + 
  28\,{{a_{1}}}^4\,{{a_{1}^{*}}}^6\,d\,{d^{*}}\,
   {f^{*}}\,{{\Sigma_{-}}}^2\,{\Sigma_{+}} + 
  2\,{{a_{1}}}^4\,{{a_{1}^{*}}}^6\,d^2\,{g'}\,
   {{\Sigma_{-}}}^2\,{\Sigma_{+}} + 
  12\,{{a_{1}}}^2\,{{a_{1}^{*}}}^8\,{{d^{*}}}^2\,
   {g'}\,{{\Sigma_{-}}}^2\,{\Sigma_{+}} + 
  3\,{{a_{1}}}^6\,{{a_{1}^{*}}}^4\,d^2\,{{g'}^{*}}\,
   {{\Sigma_{-}}}^2\,{\Sigma_{+}} + 
  8\,{{a_{1}}}^4\,{{a_{1}^{*}}}^6\,{{d^{*}}}^2\,
   {{g'}^{*}}\,{{\Sigma_{-}}}^2\,{\Sigma_{+}} + 
  6\,{{a_{1}}}^6\,{{a_{1}^{*}}}^4\,d\,f^2\,{f^{*}}\,
   {{\Sigma_{+}}}^2 + 12\,{{a_{1}}}^6\,{{a_{1}^{*}}}^4\,d\,
   f\,{{f^{*}}}^2\,{{\Sigma_{+}}}^2 + 
  8\,{{a_{1}}}^4\,{{a_{1}^{*}}}^6\,{d^{*}}\,f^2\,
   {g'}\,{{\Sigma_{+}}}^2 + 
  4\,{{a_{1}}}^4\,{{a_{1}^{*}}}^6\,{d^{*}}\,f\,
   {f^{*}}\,{g'}\,{{\Sigma_{+}}}^2 + 
  2\,{{a_{1}}}^4\,{{a_{1}^{*}}}^6\,d\,f\,{{g'}}^2\,
   {{\Sigma_{+}}}^2 + 4\,{{a_{1}}}^4\,{{a_{1}^{*}}}^6\,d\,
   {f^{*}}\,{{g'}}^2\,{{\Sigma_{+}}}^2 + 
  12\,{{a_{1}}}^6\,{{a_{1}^{*}}}^4\,{d^{*}}\,f^2\,
   {{g'}^{*}}\,{{\Sigma_{+}}}^2 + 
  6\,{{a_{1}}}^6\,{{a_{1}^{*}}}^4\,{d^{*}}\,f\,
   {f^{*}}\,{{g'}^{*}}\,{{\Sigma_{+}}}^2 + 
  6\,{{a_{1}}}^6\,{{a_{1}^{*}}}^4\,d\,f\,{g'}\,
   {{g'}^{*}}\,{{\Sigma_{+}}}^2 + 
  12\,{{a_{1}}}^6\,{{a_{1}^{*}}}^4\,d\,{f^{*}}\,
   {g'}\,{{g'}^{*}}\,{{\Sigma_{+}}}^2 + 
  12\,{{a_{1}}}^8\,{{a_{1}^{*}}}^2\,d\,f\,{{{g'}^{*}}}^2\,
   {{\Sigma_{+}}}^2 + 24\,{{a_{1}}}^8\,{{a_{1}^{*}}}^2\,d\,
   {f^{*}}\,{{{g'}^{*}}}^2\,{{\Sigma_{+}}}^2 + 
  3\,{{a_{1}}}^6\,{{a_{1}^{*}}}^4\,d^2\,f\,{\Sigma_{-}}\,
   {{\Sigma_{+}}}^2 + 8\,{{a_{1}}}^4\,{{a_{1}^{*}}}^6\,
   {{d^{*}}}^2\,f\,{\Sigma_{-}}\,{{\Sigma_{+}}}^2 + 
  12\,{{a_{1}}}^6\,{{a_{1}^{*}}}^4\,d^2\,{f^{*}}\,
   {\Sigma_{-}}\,{{\Sigma_{+}}}^2 + 
  {{a_{1}}}^4\,{{a_{1}^{*}}}^6\,{{d^{*}}}^2\,
   {f^{*}}\,{\Sigma_{-}}\,{{\Sigma_{+}}}^2 + 
  18\,{{a_{1}}}^4\,{{a_{1}^{*}}}^6\,d\,{d^{*}}\,
   {g'}\,{\Sigma_{-}}\,{{\Sigma_{+}}}^2 + 
  27\,{{a_{1}}}^6\,{{a_{1}^{*}}}^4\,d\,{d^{*}}\,
   {{g'}^{*}}\,{\Sigma_{-}}\,{{\Sigma_{+}}}^2 + 
  9\,{{a_{1}}}^6\,{{a_{1}^{*}}}^4\,d\,{d^{*}}\,f\,
   {{\Sigma_{+}}}^3 + 2\,{{a_{1}}}^6\,{{a_{1}^{*}}}^4\,d\,
   {d^{*}}\,{f^{*}}\,{{\Sigma_{+}}}^3 + 
  2\,{{a_{1}}}^6\,{{a_{1}^{*}}}^4\,d^2\,{g'}\,
   {{\Sigma_{+}}}^3 + 8\,{{a_{1}}}^8\,{{a_{1}^{*}}}^2\,d^2\,
   {{g'}^{*}}\,{{\Sigma_{+}}}^3$.

\baselineskip=\normalbaselineskip\normalsize

\vspace{5mm}
These polynomials were found using a system of analytical computations
{\it Mathematica} \cite{Mathematica}.
}\twocolumn[
\vsp\paragraph*{Appendix 1:} the polynomials, entering
to the definition of Groebner's basis (\ref{groeb}) of the system 
(\ref{maineqs1}):

\vspace{5mm}

\vspace{5mm}
The polynomials, entering to the expansion (\ref{a2cl})
of variable $a_{2}|_{S_{\pm}=0}$ in the vicinity of straight-line
string solution (\ref{1mode}):

\vspace{5mm}

\vspace{5mm}
These polynomials were found using a system of analytical computations
{\it Mathematica} \cite{Mathematica}.
]

~\newpage~\newpage
\vsp\paragraph*{Appendix 2:} ~\\
here we provide the proofs for lemmas stated above.

\vspace{2mm}\noindent 1. 
Due to implicit function theorem, it's sufficient to prove that
Jacobian of the system (\ref{maineqs}) with respect to a set
of variables $(a_{s},a_{s}^{*},\lambda)$ does not vanish
on the straight-line string (\ref{1mode}). Here $s>1$.
Computing the Jacobian, we have $\det J=2\lambda(|f|^{2}-|g|^{2})$.
Substituting the straight-line string solution (\ref{1mode}) 
to the definitions (\ref{defelem}), we have: $\lambda=|a_{1}|$,
$f=0$, $g=\{a_{1}^{2}$ at $s=2$, $0$ at $s>2\}$.
Therefore, $\det J=\{-2|a_{1}|^{3}\neq0$ at $s=2$, 
$0$ at $s>2\}$.

\vspace{2mm}\noindent 2. 
Let $s=2$. In the vicinity of straight-line string
we replace $g\to a_{1}^{2}+\epsilon g',f\to\epsilon f,d\to\epsilon d,
\Sigma_{\pm}\to\epsilon\Sigma_{\pm}$. Here we parameterize the deviation
of coefficients from straight-line string values by small parameter 
$\epsilon$ and use it as correction order counter. Let's prove that
\begin{eqnarray}
&&a_{2}|_{S_{\pm}=0}=\sum\limits_{n\geq1}{\epsilon^{n}P_{n}a_{1}^{2}
\over a_{1}^{*2n}a_{1}^{2n}},\label{a2cleps}
\end{eqnarray}
The first two equations in (\ref{maineqs}) with $S_{\pm}=0$
can be rewritten as
$$a_{2}=-{\epsilon\over a_{1}^{*2}a_{1}^{2}}\cdot a_{1}^{2}\cdot 
(\Sigma_{+}+f^{*}a_{2}^{*}+{g'}^{*}a_{2}+d^{*}a_{2}^{*2}+
\half d a_{2}a_{2}^{*}),$$
and it's complex conjugate. Here we have collected
all $\epsilon$-containing terms at the right hand side. 
Introducing variable $\tilde\epsilon=\epsilon/(a_{1}^{*2}a_{1}^{2})$
and substituting the expansion (\ref{a2cleps}), we obtain

\begin{eqnarray}
&&\sum_{n\geq1}\tilde\epsilon^{n}P_{n}=-\tilde\epsilon\Sigma_{+}
-f^{*}\sum_{n\geq1}\tilde\epsilon^{n+1}P_{n}^{*}a_{1}^{*2}\nn\\
&&-{g'}^{*}\sum_{n\geq1}\tilde\epsilon^{n+1}P_{n}a_{1}^{2}
-d^{*}\sum_{n,m\geq1}\tilde\epsilon^{n+m+1}P_{n}^{*}P_{m}^{*}a_{1}^{*4}\nn\\
&&-\half d\sum_{n,m\geq1}\tilde\epsilon^{n+m+1}P_{n}P_{m}^{*}a_{1}^{*2}
a_{1}^{2}.\nn
\end{eqnarray}

\noindent
From here we have the following recurrent relations for $P_{n}$:

\begin{eqnarray}
&&P_{1}=-\Sigma_{+},\nn\\
&&P_{n}=-f^{*}a_{1}^{*2}P_{n-1}^{*}-{g'}^{*}a_{1}^{2}P_{n-1}\nn\\
&&-d^{*}a_{1}^{*4}\sum_{1\leq m\leq n-2}P_{n-m-1}^{*}P_{m}^{*}\nn\\
&&-\half d a_{1}^{*2}a_{1}^{2}\sum_{1\leq m\leq n-2}P_{n-m-1}P_{m}^{*},
\ n>1.\nn
\end{eqnarray}

The properties of $P_{n}$ with respect to $R_{3},D_{4}$-transformations
follow from the transformation laws (\ref{astransf}) for $a_{2}$,
while the other properties of $P_{n}$ can be easily proven
from their recurrent definition by induction.

\vspace{2mm}\noindent 3. 
Expression (\ref{oint}) follows immediately from lcg-pa\-ra\-me\-tri\-za\-ti\-on
(\ref{lcgdef}) and definition of lcg-oscillators given by
formula (5) of Part I. Constraint $\chi_{3}$ generates
rotation of oscillator variables in plane $(a_{n1},a_{n2})$ and
rotation of vectors $\vec e_{1},\vec e_{2}$ about $\vec e_{3}$
in opposite direction, so that the vector field $\vec a_{n\perp}$
is preserved. $\chi_{1,2}$ generate rotations of
$\vec e_{3}$ and transformations of $\vec a_{n\perp}$
by formula (\ref{oint}). In this formula 
the curve $\vec Q$ as geometrical image is preserved,
only its parametrization is changed according to (\ref{lcgdef}).
Components of $\vec q$ are proportional to gauge fixing conditions 
(\ref{gauge}): $q_{s1}=\mbox{Re~}a_{s1}=(a_{s}+a_{-s}+a_{s}^{*}+a_{-s}^{*})/4$,
$q_{s2}=\mbox{Re~}a_{s2}=i(a_{s}+a_{-s}-a_{s}^{*}-a_{-s}^{*})/4$.
Evolution, described by $E_{0}$-transformation:
${\vec a}_{n\perp}\to{\vec a}_{n\perp}e^{-in\tau}$, is equivalent
to a motion of marked point $O$ along the curve $\vec Q(\s)$, 
uniform with respect to parameter (\ref{lcgdef}): 
$\vec Q(\s)\to\vec Q(\s+\tau)$, or in terms of $(\vec q,\vec p)$:

\begin{eqnarray}
&&\left(\begin{array}{c}\vec q\\ \vec p\end{array}\right)\to
\left(\begin{array}{cc}\cos s\tau&\sin s\tau\\ 
-\sin s\tau&\cos s\tau\end{array}\right)
\left(\begin{array}{c}\vec q\\ \vec p\end{array}\right)\nn
\end{eqnarray}
vectors $(\vec q,\vec p)$ move along ellipses in tangent planes to the sphere,
see \fref{qpfields}. In such evolution the points $\vec q=0$
move along definite curves on the sphere. On such curves the vectors 
$\vec q,\vec p$ are linearly dependent (corresponding
ellipses degenerate to segments): $\vec q\times\vec p=0$.
The vector $\vec q\times\vec p\sim i{\vec a}_{s\perp}
\times{\vec a}_{s\perp}^{*}$ is $\tau$-independent and 
always parallel to $\vec e_{3}$. Thus, orbits of $\vec q=0$ are defined
as zero-level curves of a single scalar time-independent function 
on the sphere: $F(\vec e_{3})=(\vec q\times\vec p,\vec e_{3})=0$.

\vspace{2mm}\noindent 4. 
During the period $\Delta\tau=2\pi$ all vector fields ${\vec a}_{n\perp}$
are returned to the initial state. The field $\vec a_{s\perp}$ 
at $\Delta\tau_{k}^{+}=\pi 2k/s$, $k=0...s-1$ 
is returned to the initial state, and at $\Delta\tau_{k}^{-}=\pi(2k+1)/s$, 
$k=0...s-1$ reverses the sign: $\vec a_{s\perp}\to-\vec a_{s\perp}$.
Singular points $\vec q=\mbox{Re~}\vec a_{s\perp}=0$ are preserved
by these transformations, while the points on the lifting generally
are not. 

\vspace{2mm}\noindent 5. 
For straight-line string $\vec Q(\s)\sim(\cos\s,\sin\s,0)$,
substituting it to (\ref{oint}) we have:
\begin{eqnarray}
&&~\hspace{-8mm}
\vec a_{s}\sim\oint d\s(-\sin\s,\cos\s,0)
\exp is(\s-V_{x}\cos\s-V_{y}\sin\s),\nn
\end{eqnarray}
where $\vec V=\vec e_{3}$. We see that this expression 
does not depend on $V_{z}$ and gives the same vector fields
in the vicinities of northern and southern poles.
For small $V_{x,y}$ we have in xy-plane, 
omitting unessential constant factors: 
$\vec a_{1\perp}\sim(-i,1)+o(1)$, $\vec q_{1\perp}\sim(0,1)+o(1)$
$\vec p_{1\perp}\sim(-1,0)+o(1)$ -- no singularity;
$\vec a_{2\perp}\sim(-V_{x}-iV_{y},-iV_{x}+V_{y})+o(V_{x,y})$, 
$\vec q_{2\perp}\sim(-V_{x},V_{y})+o(V_{x,y})$, 
$\vec p_{2\perp}\sim(-V_{y},-V_{x})+o(V_{x,y})$ --
saddle point; at $s>2$ $\vec a_{s\perp},\vec q_{s\perp},\vec p_{s\perp}
\sim o(V_{x,y})$ -- higher order singularity.
The evolution consists in rotation of vector fields in 
xy-plane, preserving their pole singularities.

\vspace{2mm}\noindent 6. 
The first statement follows from non-degeneracy of
saddle point and general transversality theorems 
\cite{Arnold-lectures}. 
It also can be proven by direct computation:
let's consider small deviation of solution from 
straight-line string (\ref{1mode}) and associated variations 
of the vector fields at $s=2$: 
$\vec a=(-V_{x}-iV_{y},-iV_{x}+V_{y})+(\delta a_{x},\delta a_{y})$,
$\vec q=(-V_{x},V_{y})+(\delta q_{x},\delta q_{y})$,
$\vec p=(-V_{y},-V_{x})+(\delta p_{x},\delta p_{y})$.
New singular point $\vec q=0$ will be slightly shifted
from the pole: $V_{x}=\delta q_{x}$, $V_{y}=-\delta q_{y}$
(as a result of non-degeneracy of $\df q_{x,y}/\df V_{x,y}$).
Considering the evolution $\vec q\to\vec q\cos2\tau+\vec p\sin2\tau$,
we have $V_{x}=V_{x0}+b\cos4\tau+d\sin4\tau$,
$V_{y}=V_{y0}+b\sin4\tau-d\cos4\tau$,
where $V_{x0}=(\delta q_{x}+\delta p_{y})/2$,
$V_{y0}=(-\delta q_{y}+\delta p_{x})/2$,
$b=(\delta q_{x}-\delta p_{y})/2$,
$d=(\delta q_{y}+\delta p_{x})/2$,
i.e. the saddle point moves along a circle near the pole. 
The same orbit is defined as zero level of
$(\vec q\times\vec p,\vec V)=
(V_{x}-V_{x0})^{2}+(V_{y}-V_{y0})^{2}-b^{2}-d^2$.

\vspace{2mm}\noindent 7. 
From lemma 2 we know, that each monomial in $P_{n}$
contains at least one operator $\Sigma_{(\pm)}$, additionally
$n-1$ operators from the group ($\Sigma,d,f,g'$ and their conjugates)
and $2(n-1)$ operators $(a_{1},a_{1}^{+})$.
The numbers of $\Sigma$- and $d$-operators are related: $n(\Sigma)=n(d)+1$.
Then we substitute (\ref{defelem}) to the given monomial, 
expand it to a number of (secondary) monomials and normally order each one.
Let $n(d_{\uparrow}),n(f_{\uparrow}),n(g'_{\uparrow}),n(a_{1\uparrow})$ 
be the numbers of $L_{0}^{(2)}$-raising operators in the secondary monomial,
$n(d_{\downarrow}),n(f_{\downarrow}),n(g'_{\downarrow}),n(a_{1\downarrow})$ 
be the numbers of $L_{0}^{(2)}$-lowering operators. 
Using the fact, that $d$-operators shift $L_{0}^{(2)}$ by 4, 
$f,g'$-operators -- by 2, $a_{1}^{(+)}$ by 1 (see Table~1), 
we conclude that in the matrix $(:P_{n}:)_{N_{1}N_{2}}=
\bra{L_{0}^{(2)}=N_{1}} :P_{n}: \ket{L_{0}^{(2)}=N_{2}}$
the given monomial creates non-zero block with offsets 
$\Delta_{\uparrow}L_{0}^{(2)}=\mbox{$4n(d_{\uparrow})$}
+\mbox{$2n(f_{\uparrow})$}+\mbox{$2n(g'_{\uparrow})$}
+n(a_{1\uparrow})$,
$\Delta_{\downarrow}L_{0}^{(2)}=
\mbox{$4n(d_{\downarrow})$}+\mbox{$2n(f_{\downarrow})$}+
\mbox{$2n(g'_{\downarrow})$}+n(a_{1\downarrow})$~:

\begin{figure}\label{triblocks}
\begin{center}
~\epsfxsize=6cm\epsfysize=3cm\epsffile{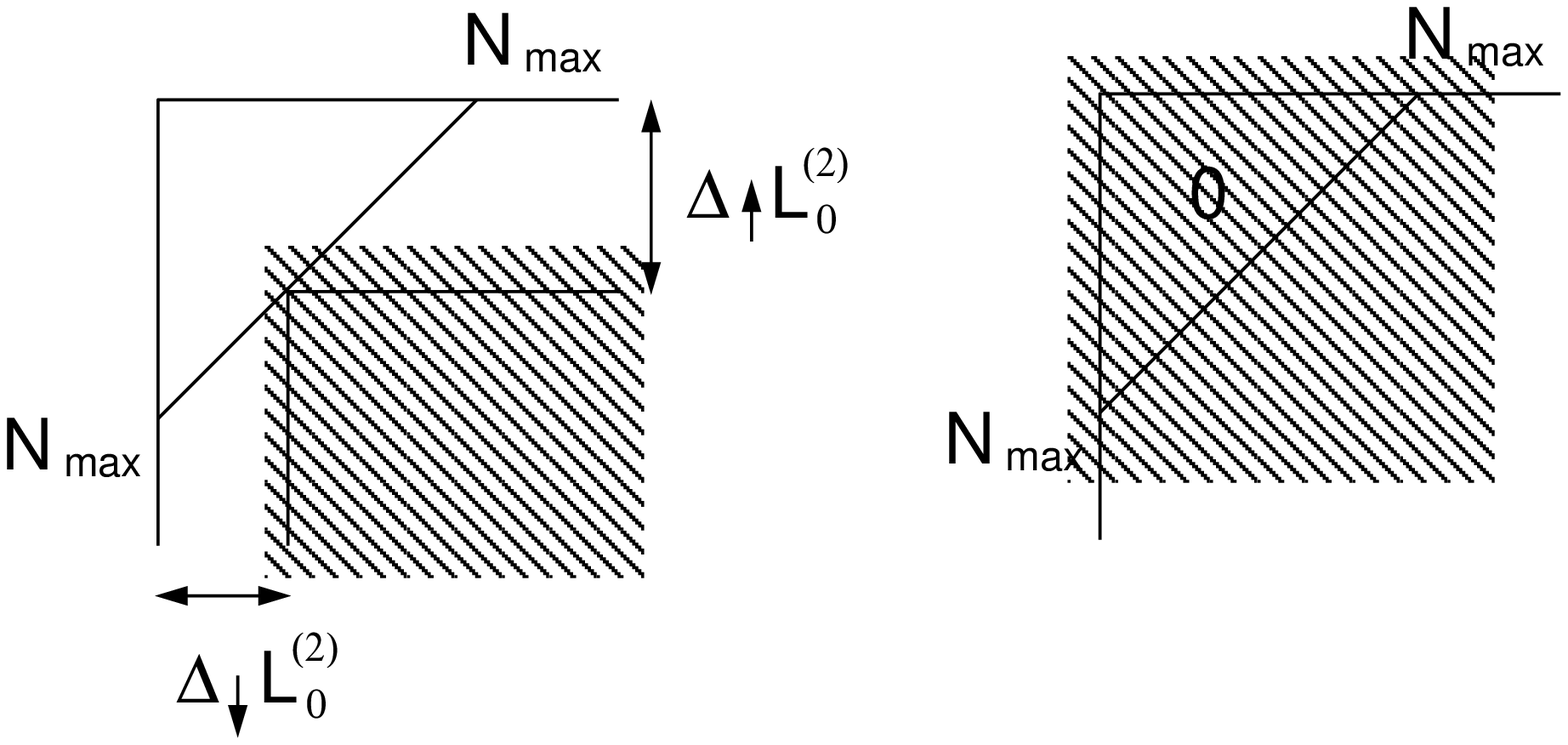}

\end{center}
\end{figure}

After summation over all secondary monomials we see that
the entries in the following triangular block vanish: 
$N1+N2<N_{max}=\Delta_{\uparrow}L_{0}^{(2)}+
\Delta_{\downarrow}L_{0}^{(2)}=4n(d)+2n(f)+2n(g')+n(a_{1})=
2n(d)+2n(\Sigma)-2+2n(f)+2n(g')+2(n-1)=4(n-1).$

\vspace{2mm}\noindent 8. 
The implication (i)$\Leftarrow$(ii) is obvious.
It's sufficient to prove (ii) for one monomial.

\footnotesize\noindent
$\bra{L_{0}^{(2)}\leq N}op_{in}...op_{i2}op_{i1}\ket{L_{0}^{(2)}\leq N}=
\sum_{Nn}...\sum_{N2}\sum_{N1}\\
\bra{L_{0}^{(2)}\leq N}op_{in}\ket{L_{0}^{(2)}=Nn}\cdot
\bra{L_{0}^{(2)}=Nn}...\ket{L_{0}^{(2)}=N2}\cdot\\
\bra{L_{0}^{(2)}=N2}op_{i2}\ket{L_{0}^{(2)}=N1}\cdot
\bra{L_{0}^{(2)}=N1}op_{i1}\ket{L_{0}^{(2)}\leq N}
$

\normalsize
\noindent
The summation here is performed over all values of $L_{0}^{(2)}$,
however, due to the $L_{0}^{(2)}$-normal ordering only
the values $L_{0}^{(2)}\leq N$ contribute, i.e. the result
can be written as a product of finite matrices
$op_{in}(N)...op_{i2}(N)op_{i1}(N)$.

\begin{figure}\label{elka}
\begin{center}
~\epsfysize=7.5cm\epsfxsize=5.3cm\epsffile{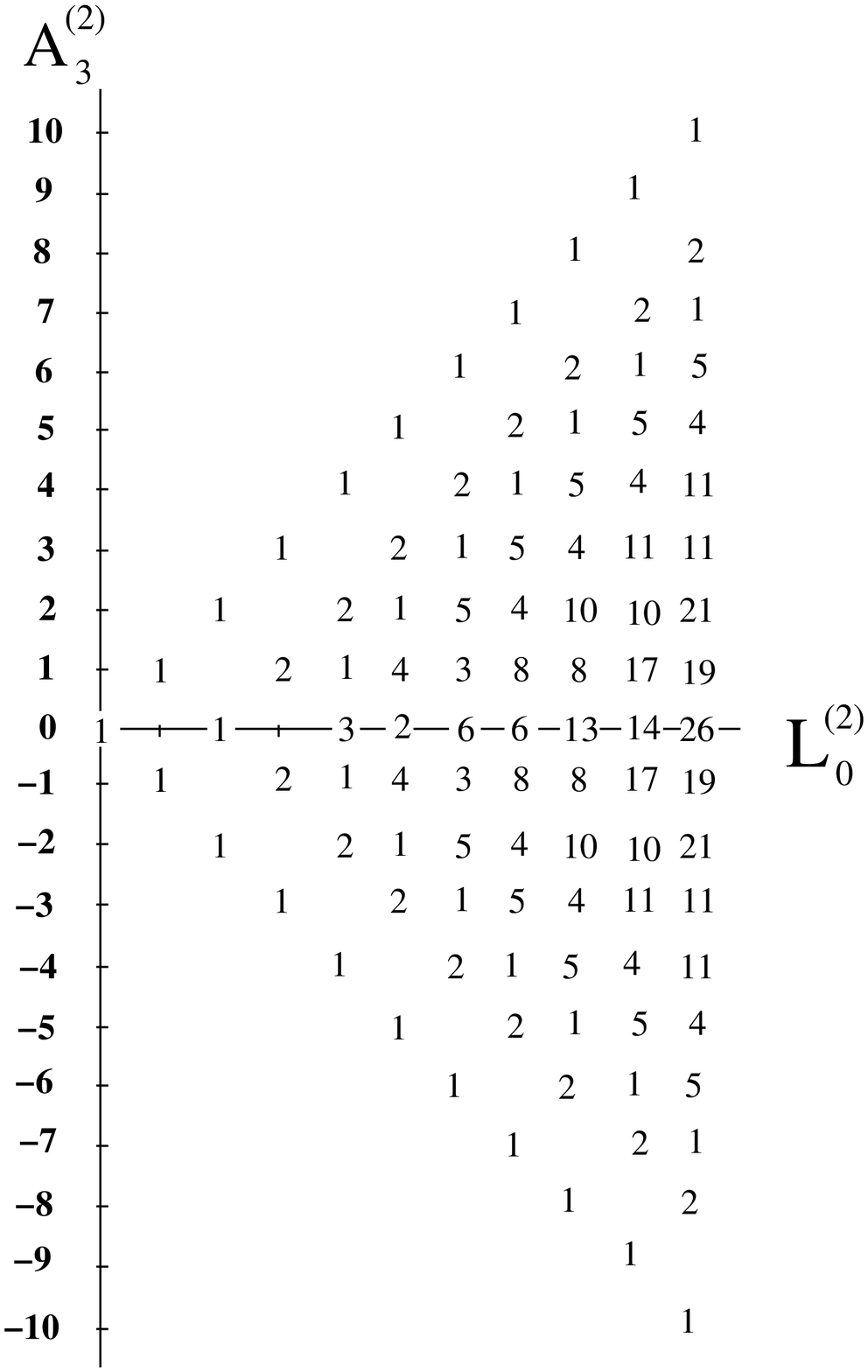}

\fignum Spectrum $(L_{0}^{(2)},A_{3}^{(2)})$.
\end{center}
\end{figure}

\vspace{2mm}\noindent 9.
The operator $a_{1}^{2}$, standing on the right of each $:P_{n}:$
in (\ref{a2q}), annulates all states of Table~2, except of $\ket{2_{1}2_{-1}}$.
This state is transferred by $a_{1}^{2}$ to $\ket{2_{-1}}$.
Each monomial of $:P_{n}:$ contains $n(\Sigma)\geq1$ operators $\Sigma_{\pm}$
and $n_{\downarrow}\geq0$ $L_{0}^{(2)}$-lowering operators from
the group $(a_{1},f_{\downarrow},g'_{\downarrow},d_{\downarrow})$,
standing on the right. Here $a_{1}$ annulates $\ket{2_{-1}}$, other
lowering operators decrease $L_{0}^{(2)}$ by 2,4 and either annulate
$\ket{2_{-1}}$ or transfer it to $\ket{0}$. The operators $\Sigma_{\pm}$
preserve $L_{0}^{(2)}$ and change $A_{3}^{(2)}$ by $\pm1$.
Figure~\ref{elka} represents the spectrum of 
$(L_{0}^{(2)},A_{3}^{(2)})$. Because there are no states immediately above
and below $L_{0}^{(2)}=A_{3}^{(2)}=0$ and $L_{0}^{(2)}=2,A_{3}^{(2)}=-1$
on this figure, $\Sigma_{\pm}$ annulate $\ket{0}$ and $\ket{2_{-1}}$.
 
\begin{figure}\label{sparse}
\begin{center}
~\epsfxsize=8cm\epsfysize=8cm\epsffile{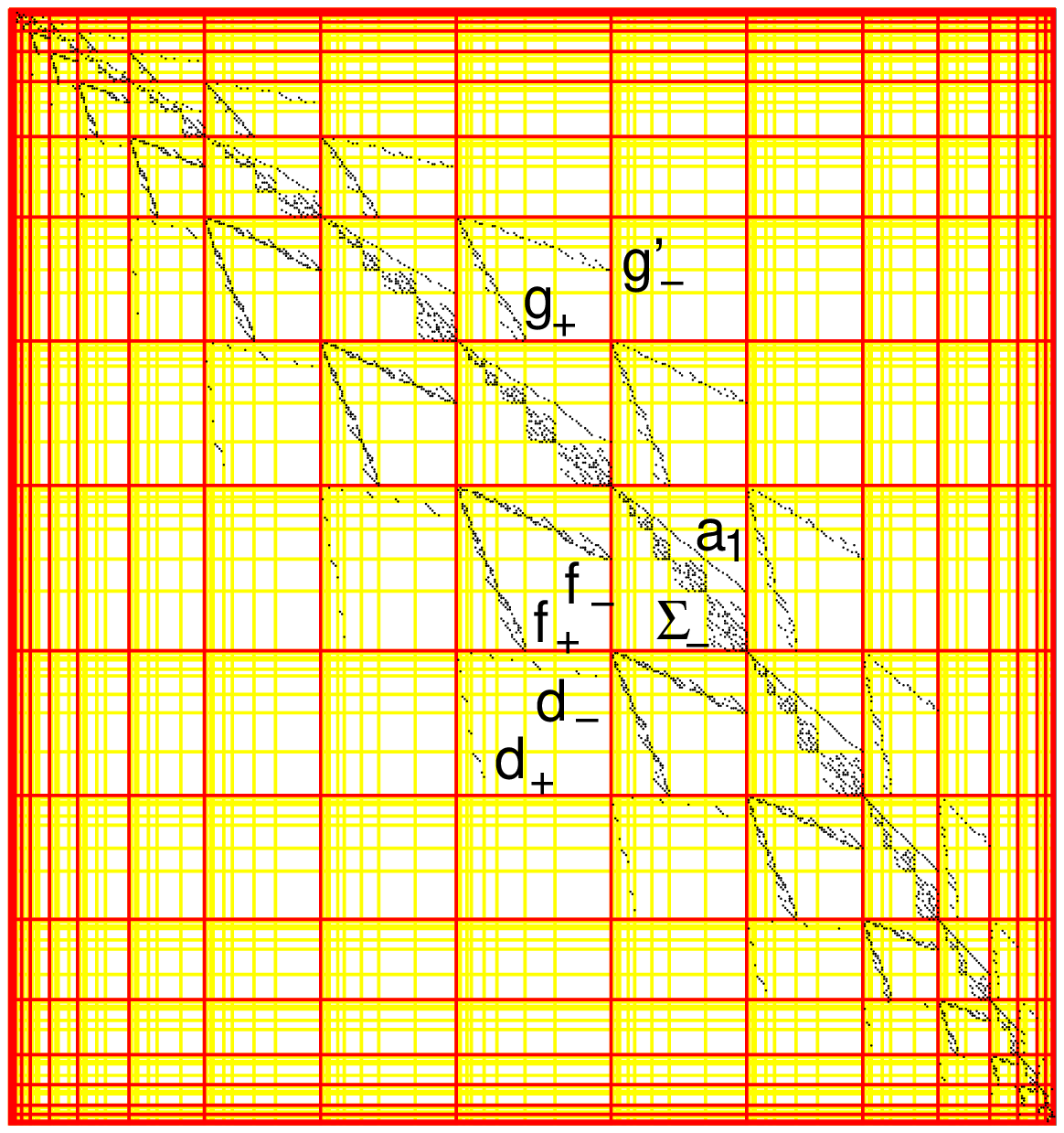}

\fignum Matrix representation of elementary operators ($N=10$).

\end{center}
\end{figure}

\vsp\paragraph*{Appendix 3:} numerical methods.\\
The matrix elements of elementary operators (Table~1) in 
$\ket{L_{0}^{(2)}\leq N}$ space are represented by sparse matrices,
where the typical pattern of non-zero entries is shown 
on \fref{sparse}. The matrices possess two-level block structure,
corresponding to their $(A_{3}^{(2)},L_{0}^{(2)})$-charge
properties. Here large blocks have constant $A_{3}^{(2)}$ values,
inside them there are smaller blocks with constant $L_{0}^{(2)}$.
There is an approximate similarity among the blocks.
Inside the blocks of the second level the matrices possess 
sparse structure of non-classified type (``other!''~\cite{numrec}).

The most intensive part of the computation is
the evaluation of large polynomials of these matrices.
This is done using sparse matrix algorithms \cite{numrec},
where non-zero elements of the matrices are stored in 
condensed form. For sparse matrix multiplication 
two algorithms of these kind are described in \cite{numrec}:
pattern multiplication and threshold multiplication. The algorithms
compute a product $C=A\cdot B^{T}$, where the transposition
is computationally expensive operation due to associated
changes of storage scheme, and is performed once
for all operators in the Table~1. The pattern multiplication algorithm
is applicable in the case when the structure of the product is known
in advance and has computational cost $T_{1}=O(N_{C}b)$, where $N_{C}$ 
is a number of non-zero entries in $C$, $b$ is an average 
number of non-zero entries per one row in $A,B$ (``bandwidth''). 
The threshold multiplication algorithm works in the case when
the structure of the product is unknown and requires 
$T_{2}=O(d^{2}b)$ operations, where $d$ is the dimension of matrices. 
Multiplication of dense matrices requires $T_{3}=O(d^{3})$ operations,
thus practically $T_{1}<\!<T_{2}<\!<T_{3}$. We have implemented a combination
of pattern and threshold multiplication algorithms, which uses known 
block structure of matrices and effectively takes into account the remaining
sparse structure inside the blocks. Separability of the computation 
to the subspaces with different values $(A_{3},Q)$ provides additional
optimization. Computing the products 
$prod_{n}=prod_{n-1}\cdot op_{i(n)}^{T}$, we initialize 
$prod_{0}$ by the projector to a given $(A_{3i},Q_{i})$-subspace,
represented as a sparse matrix. Although the intermediate products
belong to different $(A_{3},Q)$-blocks, this initialization
automatically cuts off irrelevant contributions, considerably
reducing the computational time. The localization of the final result 
in $\bra{A_{3f},Q_{f}}prod\ket{A_{3i},Q_{i}}$ block,
defined by its charge properties, is useful as consistency test. 
The resulting monomials are injected to a dense matrix $a_{2}$,
used to compute a symmetric matrix $P^{2}$ by the formula (\ref{P2q}).
Its eigenvalues are then determined by Householder algorithm 
\cite{numrec}, implemented in the library of dense matrix computations 
{\it NewMat}~\cite{newmat}. The computation is iterated for
different correction orders $n\leq nmax=6$. The scheme of main algorithm
is given on \fref{algscheme}. The necessary input data are
provided by the following external program modules: 

\begin{figure}\label{algscheme}
\begin{center}
~\epsfxsize=6cm\epsfysize=5cm\epsffile{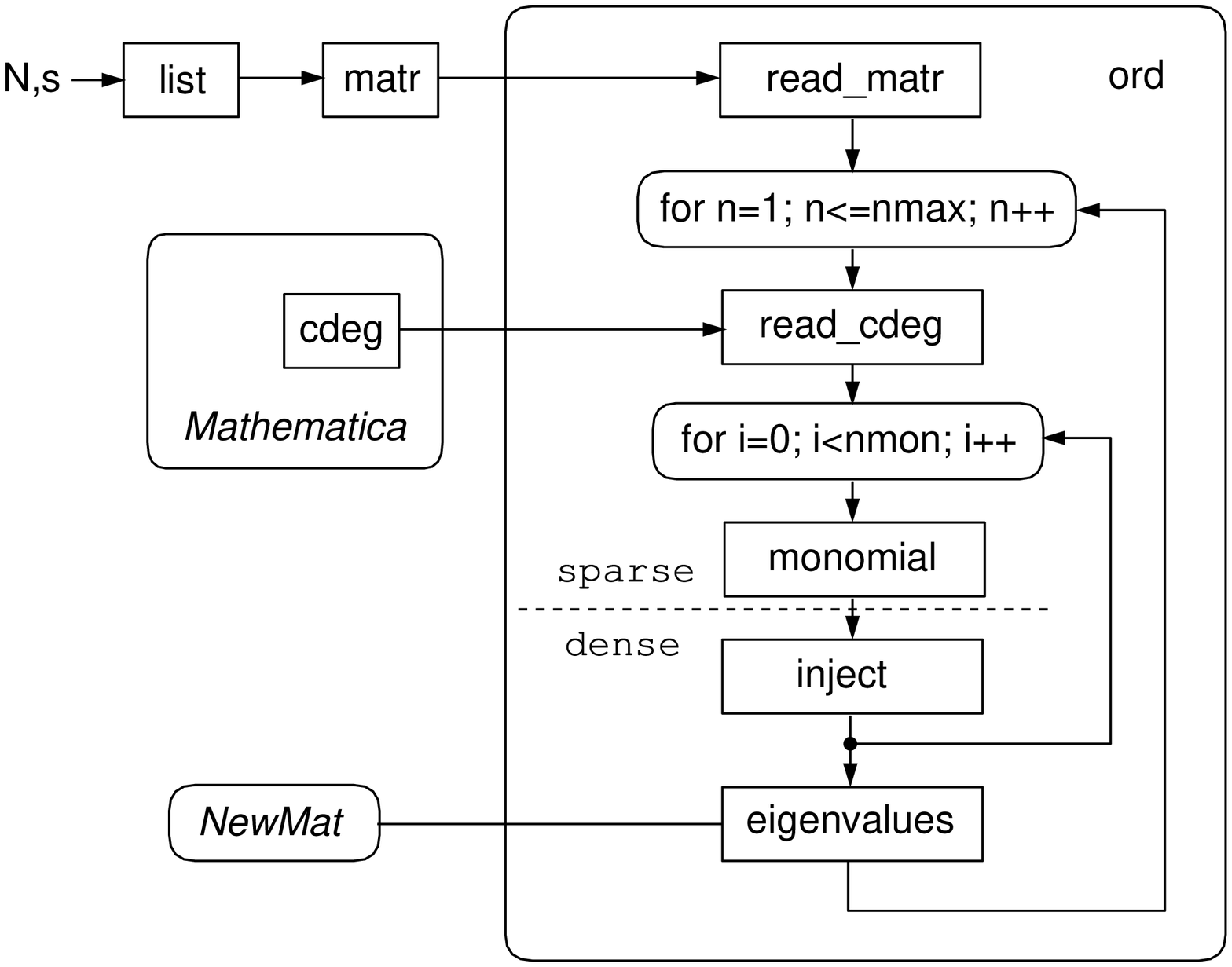}

\fignum Scheme of computation.

\end{center}
\end{figure}

\begin{itemize}
\item {\it list} -- computes for the given
$N,s$ the spectrum of $(L_{0}^{(s)},A_{3}^{(s)})$ (see \fref{elka}) 
and corresponding basis in the space $\ket{L_{0}^{(s)}\leq N}$; 
\item {\it matr} -- computes the matrix elements of the elementary operators
in this basis; 
\item {\it cdeg} -- {\it Mathematica} script, which
substitutes the definitions of elementary operators in the
polynomial expressions of Appendix~1, expands them to
normally ordered monomials and represents the result
in the form of coefficient-degree lists. The number of monomials
for each order $n=1..6$ is given in the following table:
\end{itemize}

\footnotesize

\begin{center}
\begin{tabular}{|c|cccccc|}\hline
$n$&1  &2 &3 &4 &5 &6\\\hline
num. of monomials&1&4&19&82&297&922\\\hline
\end{tabular}
\end{center}

\normalsize

\begin{figure}\label{ustakan}
\begin{center}
~\epsfxsize=8cm\epsfysize=4cm\epsffile{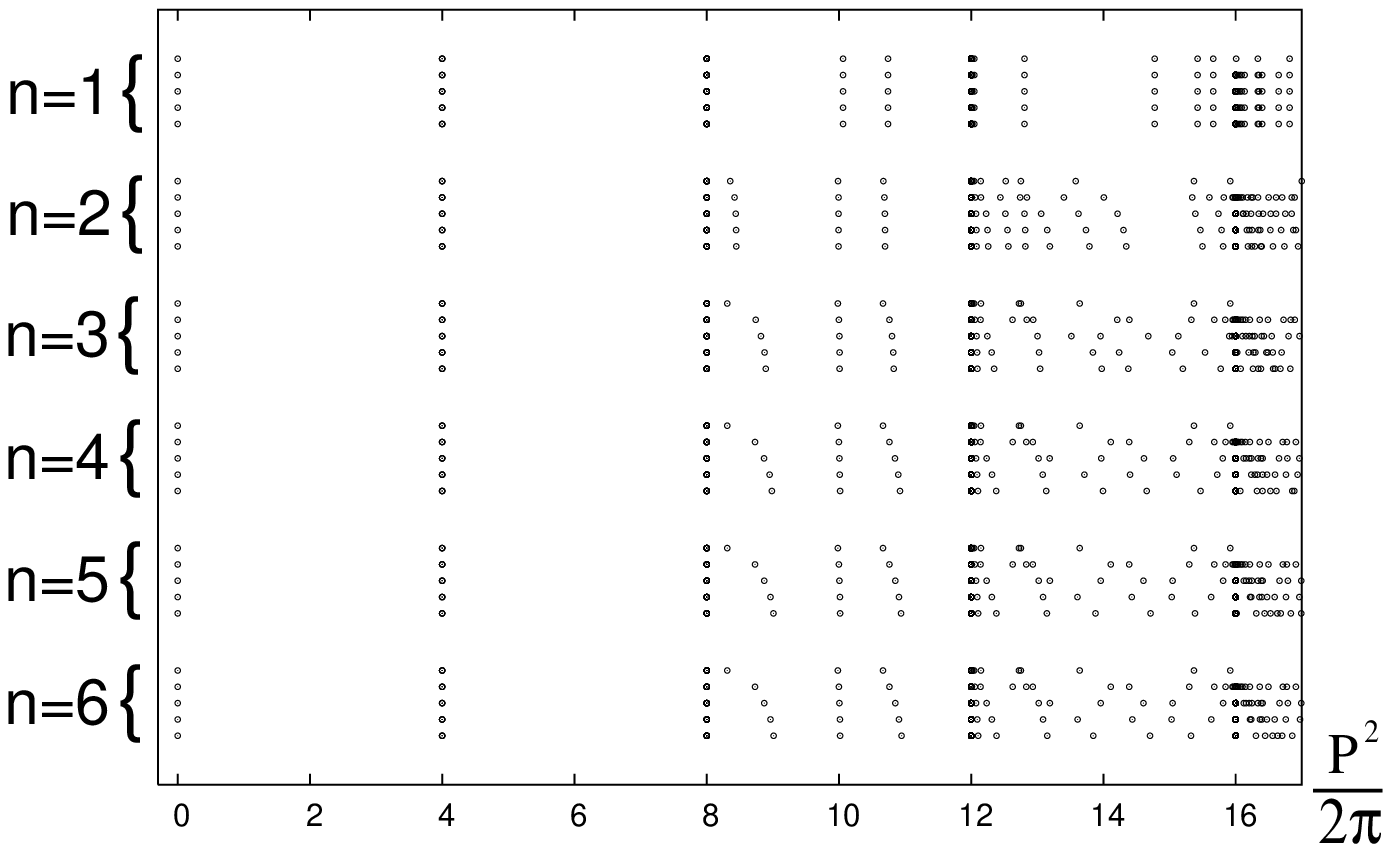}

\fignum $P^{2}/2\pi$ as a function of $n,N$-parameters.

\end{center}
\end{figure}

\vspace{3mm}
The determined spectrum of mass $(P^{2}/2\pi,S=0,Q=0)$
as a function of the correction order $n$ 
and cutoff parameter in $L_{0}^{(2)}\leq N$
is presented on \fref{ustakan}. The spectrum is subdivided
to rows with fixed $n$. The points in each row show,
in the order up-to-down, the spectra for $N=12,16,20,24,28$
(sector $Q=L_{0}^{(2)}\mod4=0$ is defined by values $N\mod4=0$).
There is no $N$-dependency in $n=1$ row, because corresponding correction 
$P_{1}=-\Sigma_{+}$ commutes with $L_{0}^{(2)}$,
thus non-zero entries of $P^{2}$ for this correction
are located in 
$\bra{A_{3}^{(2)}L_{0}^{(2)}}P^{2}\ket{A_{3}^{(2)}L_{0}^{(2)}}$
diagonal blocks. When $N$ is increasing, new blocks appear,
which give contributions to higher regions of $P^{2}$,
leaving intact already filled eigenvalues. 
Starting from $n=2$, $L_{0}^{(2)}$-nondiagonal contributions
appear, leading for some eigenvalues to $N$-dependence saturated 
at large~$N$. Other eigenvalues, as explained earlier,  
correspond to the states annulated by $a_{2}$ and remain to be constant, 
integer and degenerate. Spectra in other $Q$-sectors show similar behavior.


Time and memory requirements of the algorithm are 
presented by the following statistics:

\footnotesize 

\vspace{3mm}
\noindent
\begin{tabular}{|c|ccccc|}\hline
$N$&12  &16 &20 &24 &28\\\hline
$subdim$&69&258&890&2851&8567\\\hline
req.memory&1.4Mb&4.4Mb&23Mb&147Mb&1.1Gb\\\hline
comp.time&2~sec&44~sec&14~min&4~hours&66~hours\\\hline
\end{tabular}


\normalsize

\vspace{3mm}
Here $subdim$ is the dimension of largest $(A_{3}=0,Q=0)$ block
in dense matrices, which at large $N$ mainly defines the amount 
of required memory. The computation is performed on 
HP 2GHz Linux PC.


\begin{thebibliography}{99}
\bibitem{partI}
I. Nikitin, String theory in Lorentz-invariant light cone gauge,
hep-th/9906003
\bibitem{slstring}
E.B. Berdnikov, G.G. Nanobashvili, G.P. Pron'ko, 
The Relativistic Theory for Principal Trajectories
and Electromagnetic Transitions of Light Mesons,
Int. J. Mod. Phys. A 1993. V.8. N14. P.2447; V.8. N15. P.2551.
\bibitem{ideals}
D. Cox, J. Little, D. O'Shea, Ideals, Varieties, and Algorithms,
Springer 1998.  
\bibitem{Mathematica}
S. Wolfram, The Mathematica$^{R}$ Book, Cambridge University Press 1999.
\bibitem{DNF}
B.A. Dubrovin, S.P. Novikov, A.T. Fomenko,
Modern Geometry, Moscow, Nauka 1979.
\bibitem{Gribov}
V.N. Gribov, Quantization of non-abelian gauge theories,
Nucl.Phys. B139. 1978. P.1. 
\bibitem{Weyl}
H. Weyl, The theory of groups and quantum mechanics, New York: Dover, 1950. 
\bibitem{matrix_pol}
I. Gohberg, P. Lancaster, L. Rodman, Matrix Polynomials,
Academic Press, 1982.
\bibitem{matrix_bible}
J.H. Wilkinson, C. Reinsch, Linear Algebra, vol.2 of Handbook 
for Automatic Computations, New York: Springer Verlag 1971.
\bibitem{Arnold-lectures}
V.I. Arnold, Lectures on bifurcations and versal families,
Rus.Math.Surv. (Uspekhi Mat. Nauk) 1972. V.27 P.119.
\bibitem{numrec}
W.H. Press, S.A. Teukolsky, W.T. Vetterling, B.P. Flannery,
Numerical Recipes in C, Cambridge University Press, 1988.
\bibitem{newmat}
R. Davies, NewMat C++ Matrix Class,\\
{\tt http://ideas.uqam.ca/ideas/data/Softwares/ 
codccplusnewmat.html }
 \end{thebibliography}
\end{document}